\documentclass{aa}
\usepackage{txfonts}
\usepackage{natbib}
\usepackage{graphicx}

\bibpunct[, ]{(}{)}{;}{a}{}{,}

\def\hd{HD~101065}
\def\ps{PS}
\def\atlas{{\sc ATLAS9}}
\def\width{{\sc WIDTH9}}

\def\vald{{\sc VALD}}

\newcommand{\dream}{{\sc DREAM}}

\def\llm{{\sc LLmodels}}
\def\atl{{\sc ATLAS9}}

\def\vald{{\sc VALD}}
\def\logg{\log g}
\def\teff{T_{\rm eff}}

\def\kms{km\,s$^{-1}$}

\def\halpha{H$\alpha$}
\def\hbeta{H$\beta$}

\def\ddafit{{\sc DDAFit}}

\def\synthmag{{\sc Synthmag}}
\def\ei{$E_{\rm i}$}
\def\loggf{$\log(gf)$}

\begin{document}

\title{Realistic model atmosphere and revised abundances \\ of the coolest Ap star \hd
\thanks{Based on observations collected at the European Southern Observatory (Paranal, La Silla) and on data retrieved from the ESO Science Archive.}}

\author{D. Shulyak\inst{1,3} \and T. Ryabchikova\inst{2,3} \and R. Kildiyarova\inst{4} \and O. Kochukhov\inst{5}}
\offprints{D. Shulyak, \\
\email{denis.shulyak@gmail.com}}
\institute{Institute of Astrophysics, Georg-August-University, Friedrich-Hund-Platz 1, D-37077 G\"ottingen, Germany \and
Institute of Astronomy, Russian Academy of Science, Pyatnitskaya 48, 119017 Moscow, Russia \and
Institute of Astronomy, Vienna University, T\"urkenschanzstrasse 17, A-1180, Vienna \and
Institute of Spectroscopy, Russian Academy of Science, Physicheskaya 5, 142190 Troitsk, Russia \and
Department of Physics and Astronomy, Uppsala University, Box 515, 751 20, Uppsala, Sweden}

\date{Received / Accepted}

\abstract
{}
{Among the known Ap stars, \object{\hd} is probably one of the most interesting objects, demonstrating very rich spectra
of rare-earth elements (REE). Strongly peculiar photometric parameters of this star that can not be fully reproduced by any modern theoretical
calculations, even those accounting for realistic chemistry of its atmosphere. In this study we investigate a role
of missing REE line opacity and construct a self-consistent atmospheric model based on accurate abundance and 
chemical stratification analysis.}
{We employed the \llm\ stellar model atmosphere code together with \ddafit\ and \synthmag\ software packages to derive homogeneous and stratified
abundances for 52 chemical elements and to construct a self-consistent model of \hd\ atmosphere. The opacity
in REE lines is accounted for in details, by using up-to-date extensive theoretical calculations.}
{We show that REE elements play a key role in the radiative energy balance in the atmosphere of \hd, leading
to the strong suppression of the Balmer jump and energy redistribution very different from that of normal stars. Introducing new line lists of REEs 
allowed us to reproduce, for the first time, spectral energy distribution of \hd\ and achieve a better agreement between the unusually small observed Str\"omgren $c_{\rm 1}$ index 
and the model predictions. Using combined photometric and spectroscopic approaches and based on the iterative procedure of
abundance and stratification analysis we find effective temperature of \hd\ to be $\teff=6400$~K.}
{}


\maketitle

\section{Introduction}
Since its discovery by A.~Przybylski in 1961 \citep{przybylski} and named later after him, the
Przybylski's star (\hd, hereafter \ps) remains one of the most intriguing objects among Chemically Peculiar (CP) stars.
These objects are usually characterized by abundance anomalies in their atmospheres and by non-uniform horizontal and
vertical distributions of chemical elements and/or surface magnetic fields of different intensities.

From many observed characteristics \ps\ is a typical representative of the coolest part of 
rapidly oscillating A-type (roAp) stars.
It pulsates with the typical roAp period, $\approx$12 min \citep{kurtz79}, and it was the first roAp star discovered. It possesses a $2.3$~kG surface magnetic field \citep{cowley2000} and 
its low Fe abundance (an order of magnitude below the solar value)
follows the trend of Fe abundance versus effective temperature for Ap stars in 6300--14\,000~K range \citep{213637,cp-signature}.
A.~Przybylski was the first to note a low Fe abundance and a strong overabundance of some rare-earth (REE) elements \citep{przybylski1966}. These features were 
confirmed in many subsequent spectroscopic studies of this star \citep{wegner1974,cowley1977,cowley1998,cowley2000}. 
For instance, in the latter study \citet{cowley2000} 
reported an overabundance by up to $4$~dex (compared to the Sun) for elements heavier than Ni and enormously strong lines of the second 
ions of REEs compared to those of the first ions. This effect which is now commonly known as REE anomaly \citep{ree-anomaly,cp-signature}. 

As new 
data for REE transitions becomes available via laboratory measurements, more lines of REEs are identified and measured
in spectra of cool CP stars, thus improving abundance results. Later on it it was discovered that \ps\ is not 
even a champion in REE overabundances. At least a couple of hotter Ap stars: 
HD~170973 -- \citet{kato}; HD~144897 -- \citet{hd144897}, with accurately derived REE abundances from numerous lines of the first
and second ions have similar or even higher REE atmospheric overabundances. But a combination off a low temperature, low iron-peak
abundances and large REE overabundances result in the extremely unusual observed spectrum of \ps: weak lines of the iron-peak
elements are lost in a forest of strong and numerous REE lines. For example, the typical density of only classified spectral 
features contributing to the theoretical spectra at $5375$~\AA\ is 6--10 lines per \AA, and still not enough for proper description 
of the \ps\ spectrum since every second feature can not be accounted for in the modern spectrum synthesis.

Accurate spectroscopic analysis of \ps\ requires dedicated model atmospheres. Indeed, abnormally strong REE absorption
indicates that the atmospheric structure of the star may deviate significantly from that of normal, solar abundance stars. This is clearly seen, for instance, for the Str\"omgren $c_{\rm 1}$ photometric
index that is close to zero, which is very different than for normal stars.

Previous attempt by \citet{piskunov2001} to account for peculiar absorption in the spectrum of PS and to fit color-indices
based on the model atmosphere techniques faced a difficulty of the absence of complete line lists of REE and the impossibility
to fit simultaneously photometric and spectroscopic features (like hydrogen lines) with the same model atmosphere.
\citet{piskunov2001} used \atl\ \citep{a9-1,a9-2} atmospheres with opacity distribution functions (ODF) recalculated for individual abundances
and artificially enhanced line (but not continuum) opacity of iron to simulate the missing REE opacity. They finally adopted $\teff=6600$~K,
$\logg=4.2$ model as a preferable choice for spectroscopic analysis. However, this model failed to explain peculiar photometric
parameters of the star. Enhanced metallicity model by \citet{piskunov2001} lead to the decrease of $c_{\rm 1}$ index
but its theoretical magnitude is still too high compared with the observed value. This model was used in extensive abundance study of \ps\ by \citet{cowley2000}.

Another striking evidence of the abnormality of the atmospheric structure of \ps\ is the presence of the so-called 
core-wing anomaly in the hydrogen Balmer lines \citep{cwa-cowley}. This abrupt transition between 
the Doppler core of the hydrogen lines and their Stark wings has been modeled empirically by \citet{cwa} in terms 
of a temperature increase by 500--1000~K at intermediate atmospheric heights but so far has not been explained theoretically.

Thus, \hd\ remains an extremely challenging and very intriguing target for the application of advanced model atmosphere analysis and high-resolution
spectroscopy. In this paper we present a new attempt to combine both of these approaches and to construct a self-consistent
model atmosphere of \ps\ taking into account realistic chemistry of its atmosphere and employing up-to-date theoretical
calculations of the spectra of selected REEs.

Starting with a brief description of observations in Sect.~\ref{sec:obs}, we then
present methods of analysis in Sect.~\ref{sec:methods}
and describe the new lists of REE lines employed in the model atmosphere calculations.
Results of our analysis are presented in Sect.~\ref{sec:results} and are followed by general conclusions in Sect.~\ref{sec:conclusions}.
Our paper concludes with a brief discussion in Sect.~\ref{sec:discussion}.

\section{Observations}
\label{sec:obs}

In our study we employed several existing spectra of \ps: NTT spectrum with the resolving power $R\approx80\,000$ in the $3950-6630$~\AA\ wavelength region, 
\citep{cowley2000}, UVES spectrum in the $3100$--$10\,000$~\AA\ region with the resolving power $R\approx80\,000$ \citep{cwa,Ca-isotope}, and
an averaged UVES spectrum from the time-series observations obtained in the 4960--6990~\AA\ spectral region with the
resolving power $R\approx115\,000$ \citep{lpv07,ryabchik07}. 

Details of the spectroscopic observations and full description of the data reduction are given in the
corresponding papers. Direct comparison of the normalized spectra reveals a remarkable agreement between all observations. This allows us to
use the equivalent width measurements from \citet{cowley2000}, supplementing them with the measurements in the spectral region 
$\lambda>6600$~\AA\ using the other two spectra.         

\section{Methods}
\label{sec:methods}
\subsection{Abundance analysis}

The main goal of the present paper is to calculate a model atmosphere of \ps\ accounting for its anomalous
chemical composition. In magnetic chemically peculiar star an accurate abundance analysis may be carried out by fitting the magnetic 
synthetic spectrum to the observed line profiles. 
This procedure is very time consuming. Therefore, to simplify calculations,
we analysed the equivalent widths replacing magnetic effects by a pseudo-microturbulence. It is sufficient for line opacity calculations,
although the abundances derived using the common value of microturbulent velocity may be slightly inaccurate \cite[see discussion in][]{cowley2000}.
In present analysis we used the \citet{a9-2} \width\, code modified by V. Tsymbal \citep[see][]{widthV}.
 
The \vald\ database \citep{vald1,vald2} and the \dream\ REE line database \citep[and references therein]{dream99}, which is made accessible via the 
\vald\ extraction procedures, are our main sources for atomic parameters. For the REE, which provide the majority of lines in
the \ps\ spectrum, accurate laboratory transition probabilities measured by Wisconsin group are used: \citet{la2} -- \ion{La}{ii};
\citet{la2} -- \ion{La}{ii}; \citet{ce2} -- \ion{Ce}{ii}; \citet{nd2} -- \ion{Nd}{ii}; \citet{sm2} -- \ion{Sm}{ii}; \citet{eu2} -- \ion{Eu}{ii}; 
\citet{gd2} -- \ion{Gd}{ii}; \citet{tb2} -- \ion{Tb}{ii}; \citet{dy1-2} -- \ion{Dy}{i}/\ion{Dy}{ii}; \citet{er2} -- \ion{Er}{ii};
\citet{tm2} -- \ion{Tm}{ii}; \citet{hf2} -- \ion{Hf}{ii}. Among heavier elements Th and U are particularly interesting. For these species we used laboratory 
and calculated transition probabilities from the following papers: \citet{th2} -- \ion{Th}{ii}; \citet{th3} -- \ion{Th}{iii}; 
\citet{u2} -- \ion{U}{ii}. Transition probabilities for REEs in the second ionization stage were taken from the \dream\ database except
\ion{Pr}{iii} \citet{pr-NLTE}, \ion{Nd}{iii} \citet{hd144897}, \ion{Eu}{iii} \citet{eu3}, \ion{Tb}{iii} and \ion{Dy}{iii} (Ryabtsev,
private communication). 

Extended measurements and theoretical calculations for the REEs, Th and U justify revision of some of the 
partition functions (PF), which until now mainly came from \citet{a9-2} \atlas\ code. 
Our recalculated PFs are made available online\footnote{{\tt http://www.astro.uu.se/$^\sim$oleg/pf.html}}.  

We did not take into account the hyperfine structure (hfs) in our abundance analysis because the hfs constants are available only for a small subset of spectral lines. 
  
\subsection{Stratification analysis}

In the majority of cool Ap stars the atmospheres are not chemically homogeneous. Element stratification built up by atomic diffusion 
influences the atmospheric structure \citep{hd24712,acir,leblanc2009}. As it follows from \citet{ind}, Fe, Si, and Cr are the elements that
have the major effect on the atmospheric $T-P$ structure in the case of homogenous abundances.
The cumulative effect of element stratification on the model structure was demostrated by \citet{wade2003} and later by \citet{monin2007}
who used diffusion calculations for $39$ chemical elements simultaneously.
Thus, the stratification analysis of \ps\ was performed for four elements: Fe, Ba, Ca, and Si. 
It was not possible to carry out stratification analysis for Cr and other iron-peak elements
due to weakness of their absorption features and heavy blending by REE lines. 
The lower energy levels of practically all Cr lines used in the abundance determination belong to a narrow energy range, which
makes the Cr lines insensitive to abundance gradients.
Hence, we restricted stratification analysis to the four
elements since they are represented in \hd\ by a sufficient number of atomic lines, probing different
atmospheric layers and enabling line profile fitting.

We applied a step-function approximation of chemical stratification as implemented in \ddafit\ -- an automatic
procedure for determination of vertical abundance gradients \citep{synthmag07}.
In this routine, the vertical abundance distribution of an element
is described by four parameters: chemical abundance in the upper
atmosphere, abundance in deep layers, the position of the abundance jump
and its width. All four parameters can be optimized simultaneously with the
least-squares fitting procedure. 

Most unblended lines in the \ps\ spectrum are located in the red spectral region,
where line profiles are also substantially distorted by the Zeeman splitting due to wavelength dependence of the Zeeman effect. \ddafit\ enables 
an accurate stratification analysis of such
magnetically-splitted lines. In the search for optimal vertical distribution of chemical elements we use magnetic spectrum 
synthesis with the \synthmag\ code \citep{synthmag07}. This software represents an improved version of the program 
developed by \citet{P99}. 

A list of the lines used in stratification analysis
is given in Table~\ref{tab:strat} (Online material).  To get a better sensitivity of the fitted Ca distribution to the upper atmospheric layers, we used the IR-triplet \ion{Ca}{ii} line
at $\lambda$~8498~\AA. As shown by \citet{Ca-isotope} and \citet{cowley2009}, its core is represented entirely by the absorption of the heavy isotope $^{48}$Ca. Therefore
wavelength of the line of this isotope is given in Table~\ref{tab:strat}.   

\subsection{Calculation of model atmospheres}

To perform the model atmosphere calculations
we used the recent version of the \llm\, \citep{llm} stellar model atmosphere code. 
The code accounts for the effects of individual and stratified abundances.
The stratification of chemical elements is
an input parameter and thus does not change during the model atmosphere calculation process. 
This allows us to explore the changes in the model structure due to stratification that was inferred directly 
from observations, without modeling poorly understood processes that could be responsible for the observed inhomogeneities.

Note that such an empirical analysis of chemical element
stratification is based on the model atmosphere technique
and thus the temperature-pressure structure of the model atmosphere itself depends upon
the stratification which we want to determine. Therefore, the calculation of the model atmosphere
and the stratification (abundances) analysis are linked together
and an iterative procedure should be used in this case. It consists of repeated steps
of stratification and abundance analysis that provide an input for the calculations of model atmosphere
until atmospheric parameters ($\teff$, $\logg$, etc.) converge and theoretical observables (photometric colors, profiles of 
hydrogen lines, etc.) fit observations. We refer the reader to the recent papers \citep{hd24712,acir} for a more detailed description of this technique.

The \vald\ database \citep{vald1,vald2} was used as a main source of the atomic line data 
for computation of the line absorption coefficient. The recent VALD compilation contains
information on about $66\times10^6$ atomic transitions. Most of them come from the latest theoretical calculations performed by
R.~Kurucz\footnote{{\tt http://kurucz.harvard.edu}}.

Since a $2.3$~kG surface magnetic field of \ps\ is too weak to affect noticeably the atmospheric structure and
energy distribution \cite[see][for more details]{zeeman_paper1,zeeman_paper2} we adopted a $1$~\kms microturbulent velocity
to roughly account for the magnetic broadening and to avoid time-consuming model calculations with detailed polarized radiative transfer.

\subsection{REE line lists for model atmosphere calculation}
\label{sec:linelist}

The main sources for the REEs transition probabilities data between experimentally known energy levels
are \vald\, and \dream\, databases, hereafter referred to ``VALD data''. These data were supplemented by the transition probabilities
for the observed 
and predicted lines of \ion{Pr}{ii}-\ion{Pr}{iii} \citep{pr-NLTE}, \ion{Nd}{ii}-\ion{Nd}{iii} \citep{nd-NLTE,hd144897}, 
\ion{Eu}{iii} \citep{eu3}, \ion{Tb}{iii} and \ion{Dy}{iii} (Ryabtsev, private communication), hereafter referred to as ``ISAN data''. 
For the latter two ions transition probability calculations were based on the extended term analysis. 

Table\,\ref{tab:linestat}
demonstrates a dramatic difference between the number of observed (VALD) and predicted (ISAN) transitions for Pr and Nd,
in particular for the lines of the first ions. We are missing more than 90\%\, of the potential line absorbers.
The same situation is expected for lines of the other REEs. Taking this situation into account, we performed calculations for \ion{Sm}{ii} --
the second-abundant element after Nd in \ps. The energy levels of \ion{Sm}{ii} were calculated by the Hartree-Fock method implemented in the
\citet{cowan} code. The ground state of \ion{Sm}{ii} is the 4f$^6$6s configuration. In addition, our calculations include even configurations
4f$^6$nd (n=5,6), 4f$^6$7s, 4f$^5$5d6p, 4f$^5$6s6p, and odd configurations 4f$^6$np (n=6-8), 4f$^4$5d$^2$6p, 4f$^4$6s$^2$6p, 
4f$^5$5d$^2$, 4f$^5$5d6s. Calculations are based on the wave functions obtained by the fittings  the energy levels. As in the case 
of \ion{Pr}{ii} and \ion{Nd}{ii}, all Hartree-Fock transition integrals are scaled by a factor 0.85. In total, transition probabilities
for more than one million lines were calculated and added to the ISAN list of the REE lines.     

The recent studies by \citet{nd-NLTE,pr-NLTE} demonstrated that the line formation 
of Pr and Nd can strongly deviate from the local thermodynamic equilibrium.
For instance, the doubly ionized lines of these elements are unusually strong 
due to combined effects of stratification of these elements and departures from LTE. 
However, a precise NLTE analysis is beyond the scope of this paper and currently
could not be coupled to a model calculation. Therefore, we followed the approach outlined in \citet{hd24712} 
where authors used a simplified treatment of the REE NLTE opacity. 

Taking into account
a systematic difference in abundances derived for the first 
and second REE ions, it is essential to reduce the oscillator strengths for
the singly ionized REE lines by this difference in the model line list while using the abundances 
derived from second ions as an input for model atmosphere calculations (or vice versa). 
Obviously, the adopted reduction factors depend upon the assumed abundances of the first and second ions
and thus change slightly if the model parameters ($\teff$, $\logg$, etc.) are modified in the course of the iterative procedure of abundance analysis. The logarithmic scaling factors, $\log(N_{\rm II}/N_{\rm III})$, adopted for our final model are (in dex):
$-1.6$ (Ce), $-2.4$ (Pr), $-1$ (Nd), $-2.66$ (Tb), $-2.5$ (Dy), and $-1.64$ (Eu).
These factors were applied to all lines of the respective ions in the master line list.
Then this line list was processed by the line pre-selection procedure in the \llm\ code to
select only those lines that contribute non-negligibly to the total line opacity coefficient for a given temperature-pressure distribution.
Applying this line scaling procedure allowed us to mimic the line strengths correspond to the NLTE ionization equilibrium abundance for each REE \citep[see][for more details]{hd24712}.

\section{Results}
\label{sec:results}

\subsection{Adopted atmospheric parameters}

We started analysis with a homogeneous abundance model calculated with $\teff=6600$\,K, $\logg=4.2$, and
individual abundances taken from \citet{cowley2000}. 
Once the stratification was introduced in the model atmosphere,
the fit to the hydrogen \halpha\ line and the observed photometric parameters required decreasing of $\teff$ of
the star down to approximately $6400$\,K. In total, four iterations of the abundance analysis and the model atmosphere calculation
were performed to achieve a converged solution. The abundances from the third iteration did not introduce noticeable
changes in the atmospheric structure.

The effective temperature obtained in our study is identical to the $\teff$ obtained by \citet{cwa}, 
who fitted the \halpha\ and \hbeta\ lines using the same UVES spectra. However, Kochukhov et al. relied on less 
advanced \atl\ model atmospheres calculated using ODFs with artificially increased Fe line absorption \citep{piskunov2001} 
and also attempted to empirically adjust the $T$-$\tau$ structure of their model.

\begin{figure}
\includegraphics[width=\hsize]{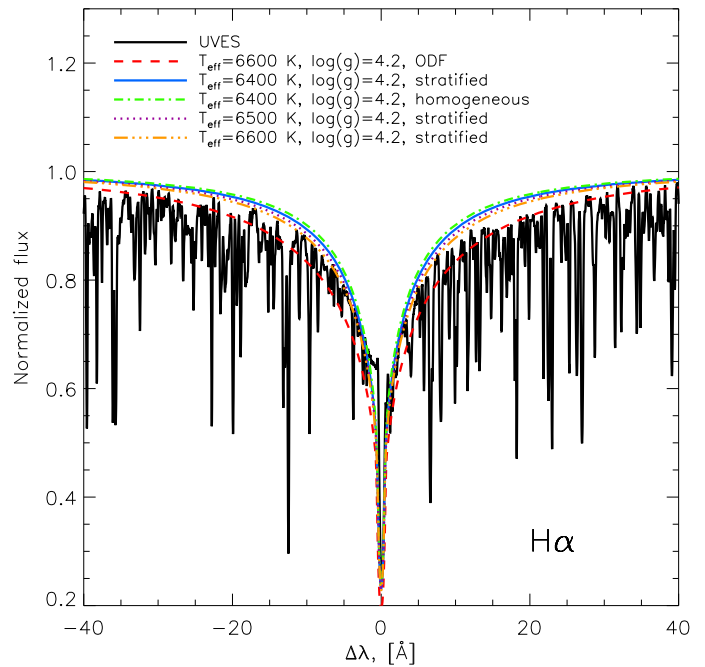}
\caption{Comparison between the observed and predicted \halpha\ line profiles.}
\label{fig:halpha}
\end{figure}

Figure~\ref{fig:halpha} illustrates the observed and predicted \halpha\ line profiles calculated with different atmospheric
models. Unfortunately, it is impossible to infer an accurate value of $\logg$ from the \halpha\ profile
due to low temperature of the star and resulting poor sensitivity of the hydrogen lines to the pressure stratification. For this reason
we kept this parameter fixed during the iterative procedure of abundance analysis (however, see Discussion).

\subsection{Abundance pattern}

Table~\ref{tab:abn} (Online material) summarizes individual abundances of  chemical elements in the atmosphere of \ps\ derived
with the final model atmosphere. Note the strong enrichment of REE elements compared to the solar composition. These
elements dominate the spectrum of \ps, playing a key role in the radiative energy balance (see below).

In Fig.~\ref{fig:abn} we illustrate abundance pattern of \ps\ relative to the recent compilation of the solar 
abundances \citep{solabn}. The elements heavier than Ba exhibit an overabundance by 3~dex and larger. 
Several REEs show clear ionization anomaly related to stratification of these elements and departures from LTE.

\begin{figure}
\includegraphics[width=\hsize]{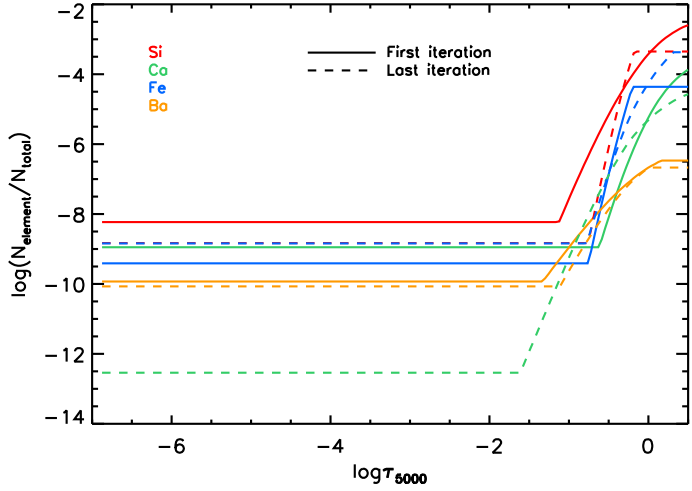}
\caption{Stratification of four elements in the atmosphere of \hd\ derived at the first and last iteration of the abundance analysis.}
\label{fig:strat}
\end{figure}

Vertical distribution of Si, Fe, Ba, and Ca is presented in Fig.~\ref{fig:strat}. This figure illustrates
the difference between the initial and final stratification as derived at the first and the last iterations respectively.
Ba seems to be the only element whose stratification
did not change appreciably during iterations. Only the position of the abundance jump changed slightly.
Fe shows a change in the abundance in the upper atmosphere by about $\sim$\,1~dex.
The position of the abundance jump for Si changed significantly.
In contrast, the Ca abundance in the upper atmosphere changes dramatically. These results illustrate the importance of taking into account a feedback of stratification on the model structure.

The effect of introducing stratification on the fit to \halpha\ line appears to be not very strong. The maximum difference
between the stratified and non-stratified model predictions is at the level of $1$\%
as seen from Fig.~\ref{fig:halpha}. However, the two profiles are still clearly distinguishable. Thus, chemical stratification should be taken into account in accurate
modeling of the hydrogen line formation. 

\begin{figure}[!t]
\includegraphics[width=\hsize]{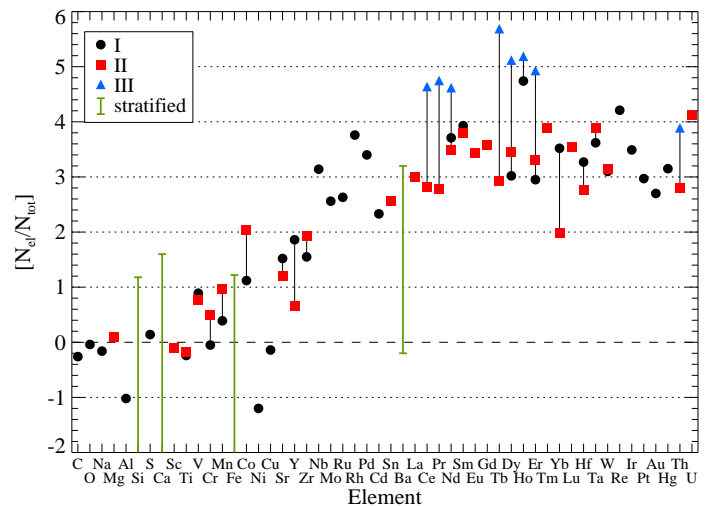}
\caption{Chemical composition of the atmosphere of \hd. Different symbols correspond to ions from I to III. Bars show abundance ranges for stratified elements.}
\label{fig:abn}
\end{figure}

The Fe and Ba abundance distributions agree with those derived previously 
by either trial-and-error method \citep{Fe-Ba} or based on the equivalent widths \citep{YGG07}.
Figures~\ref{fig:pBa}-\ref{fig:pFe} (Online material) illustrate the resulting fit to line profiles.

\onlfig{3}{
\begin{figure}
\begin{center}
\includegraphics[width=\hsize]{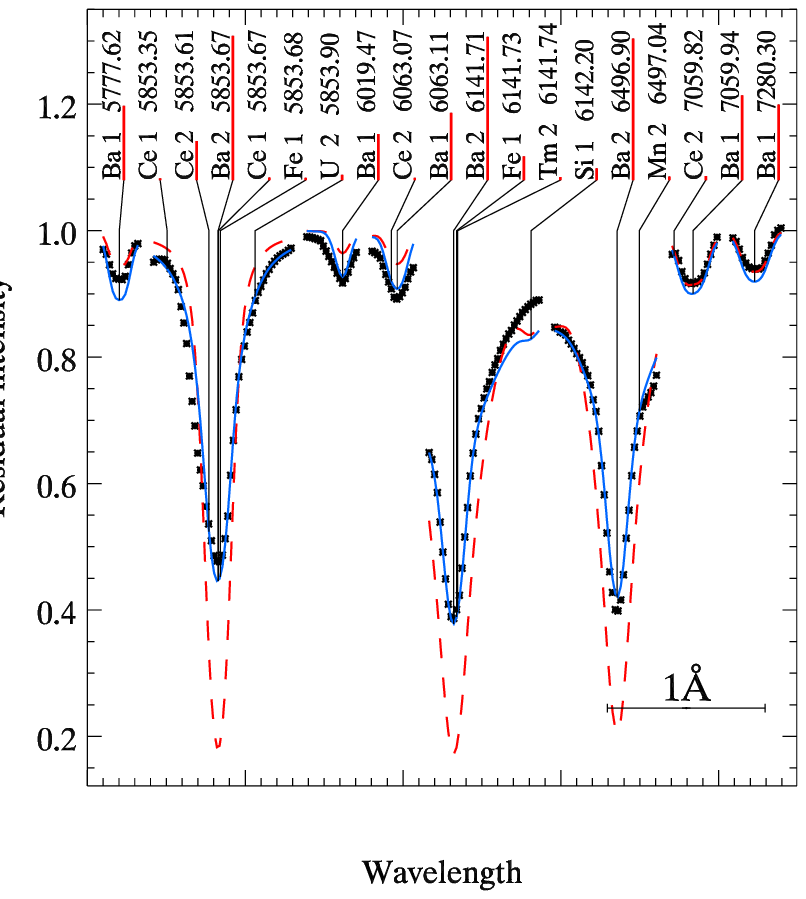}
\caption{Comparison between the observed and theoretical profiles for the Ba lines. Observations are shown with symbols, 
thick full line corresponds to the synthetic spectrum calculated with the best-fitting stratified abundance distribution 
and dashed line shows prediction of the spectrum synthesis with a homogeneous abundance.}
\label{fig:pBa}
\end{center}
\end{figure}
}

\onlfig{4}{
\begin{figure}
\begin{center}
\includegraphics[width=\hsize]{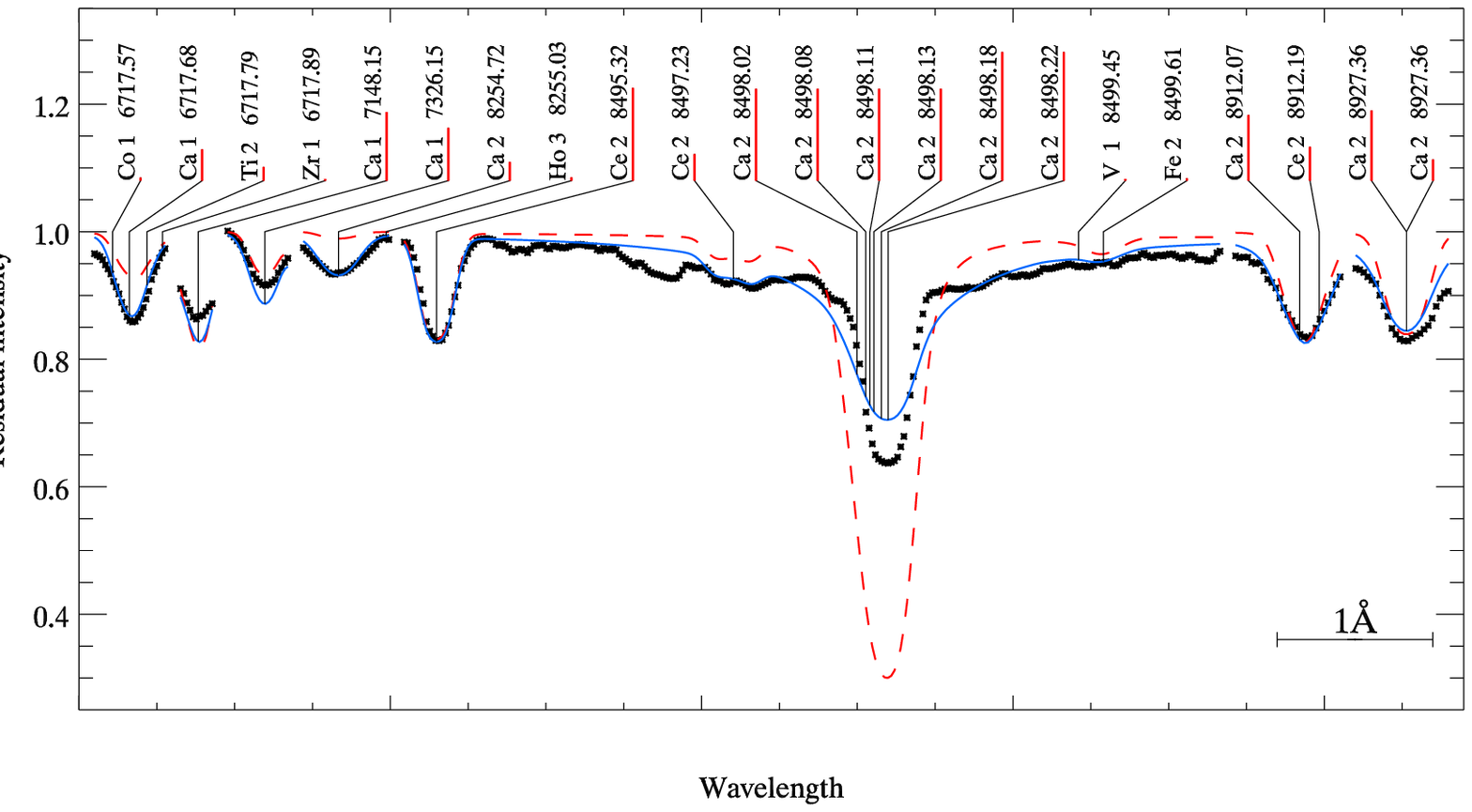}
\caption{Same as in Fig~\ref{fig:pBa} but for Ca.}
\label{fig:pCaIR}
\end{center}
\end{figure}
}

\onlfig{5}{
\begin{figure}
\begin{center}
\includegraphics[width=\hsize]{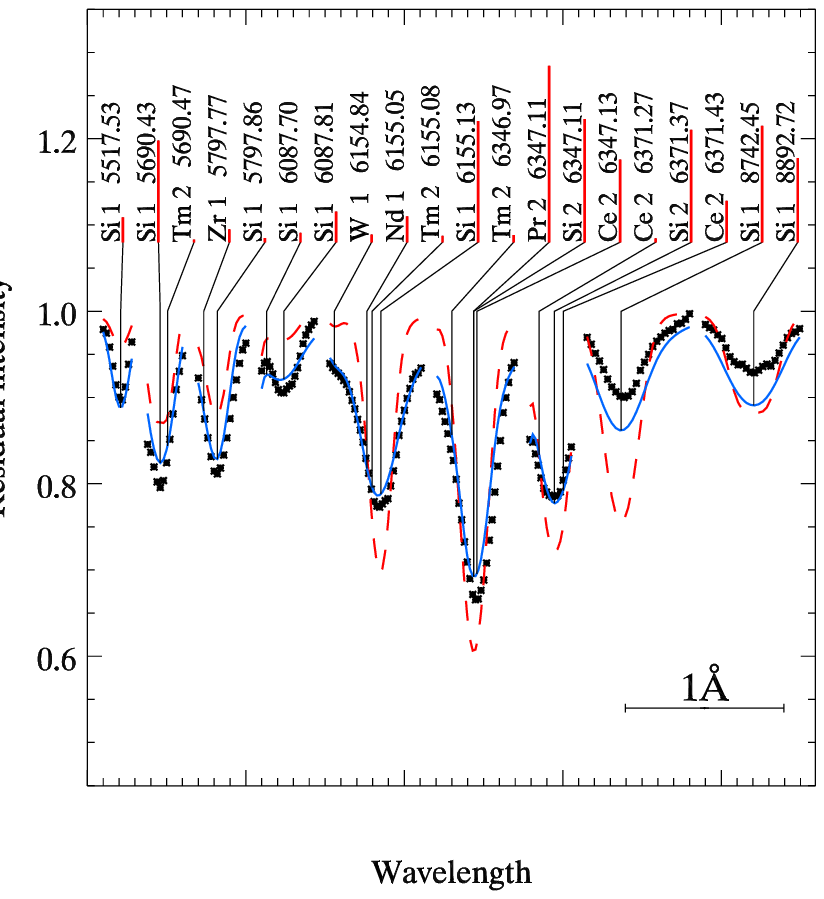}
\caption{Same as in Fig~\ref{fig:pBa} but for Si.}
\label{fig:pSi}
\end{center}
\end{figure}
}

\onlfig{6}{
\begin{figure}
\begin{center}
\includegraphics[width=\hsize]{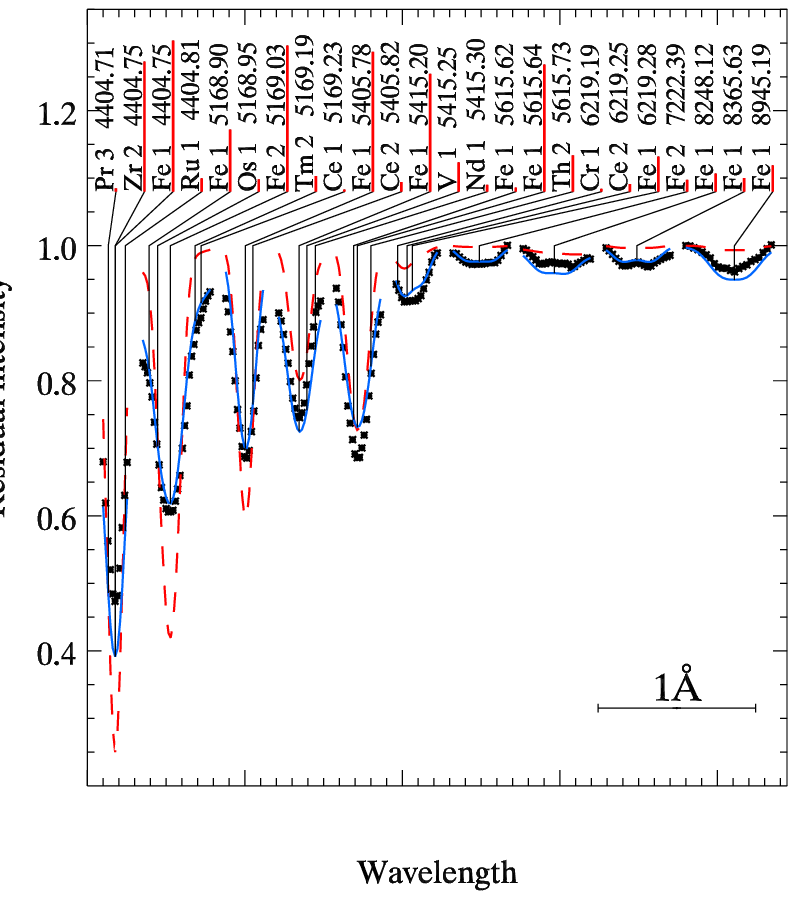}
\caption{Same as in Fig~\ref{fig:pBa} but for Fe.}
\label{fig:pFe}
\end{center}
\end{figure}
}

\subsection{Energy distribution and photometric colors}

\begin{figure*}
\includegraphics[height=\hsize,angle=90]{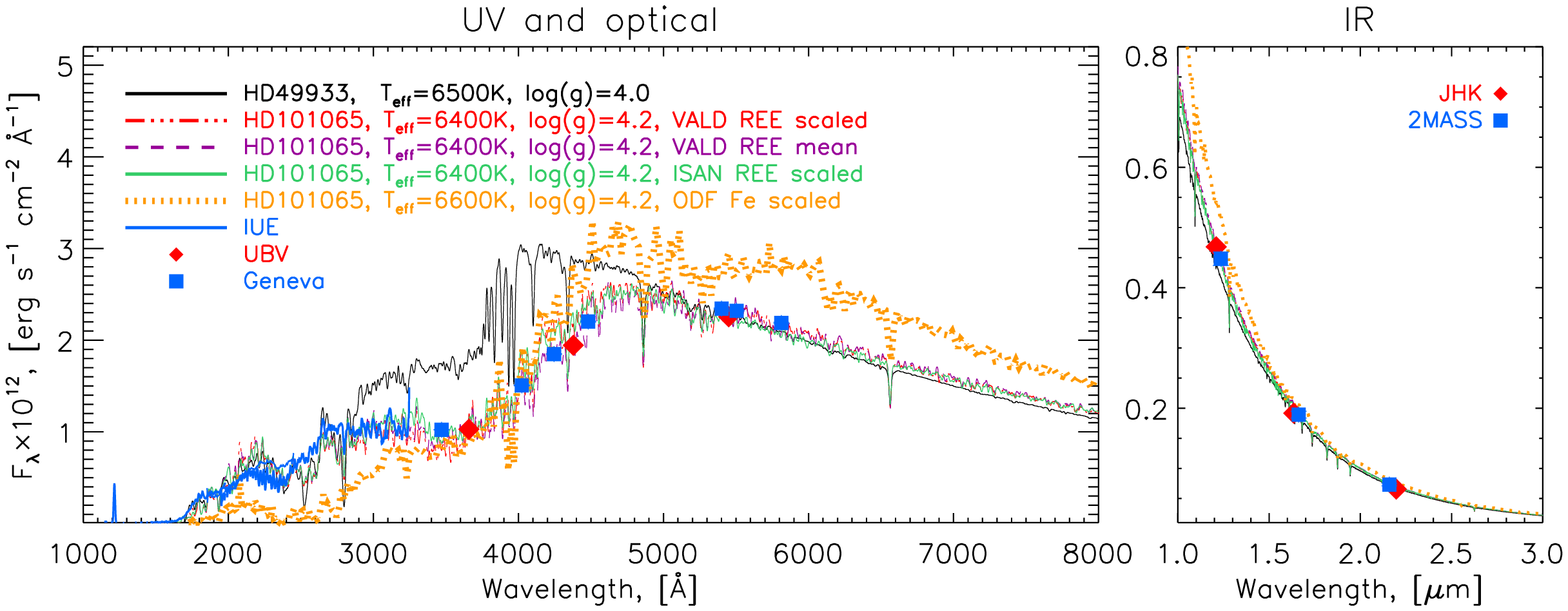}
\caption{Observed and theoretical energy distribution of \hd\ calculated using line lists from two datasets: VALD and ISAN.
For comparison purpose the theoretical energy distribution of a normal F-type star HD~49933 is also shown. 
All theoretical calculations are smoothed with $20$\AA\ Gaussian.}
\label{fig:sed}
\end{figure*}

Unusual photometric parameters of \ps\ clearly indicate that its energy distribution should differ much
from that of normal stars with similar $\teff$. Indeed, a dense forest of REE lines obtained from theoretical computations
should have a strong impact on the atmospheric energy balance. 

Figure~\ref{fig:sed} illustrates
theoretical energy distributions of \hd\ and a normal F-type star HD~49933. Both stars have close effective 
temperatures but very different atmospheric chemistry: HD~49933 has a slight underabundance of iron-peak elements
and a nearly solar abundance of REEs as derived by \citet{hd49933}. Model fluxes were computed with the stratified distribution
of Si, Ca, Fe, and Ba as well as with the scaled REE abundances as described above. In addition, \atl\ ODF model 
from \citet{piskunov2001} calculated with $\teff=6600$~K, $\logg=4.2$ and Fe opacity scaled to simulate an enhanced
REE absorption is shown for comparison purpose. To keep the figure representative, here we show only models
with final $\teff=6400$~K, $\logg=4.2$ and different assumptions about REE opacity. Energy distributions of models with 
$\teff=6500$~K and $\teff=6600$~K are displayed in Fig.~\ref{fig:sed2} (Online Material).

\onlfig{8}{
\begin{figure*}
\includegraphics[height=\hsize,angle=90]{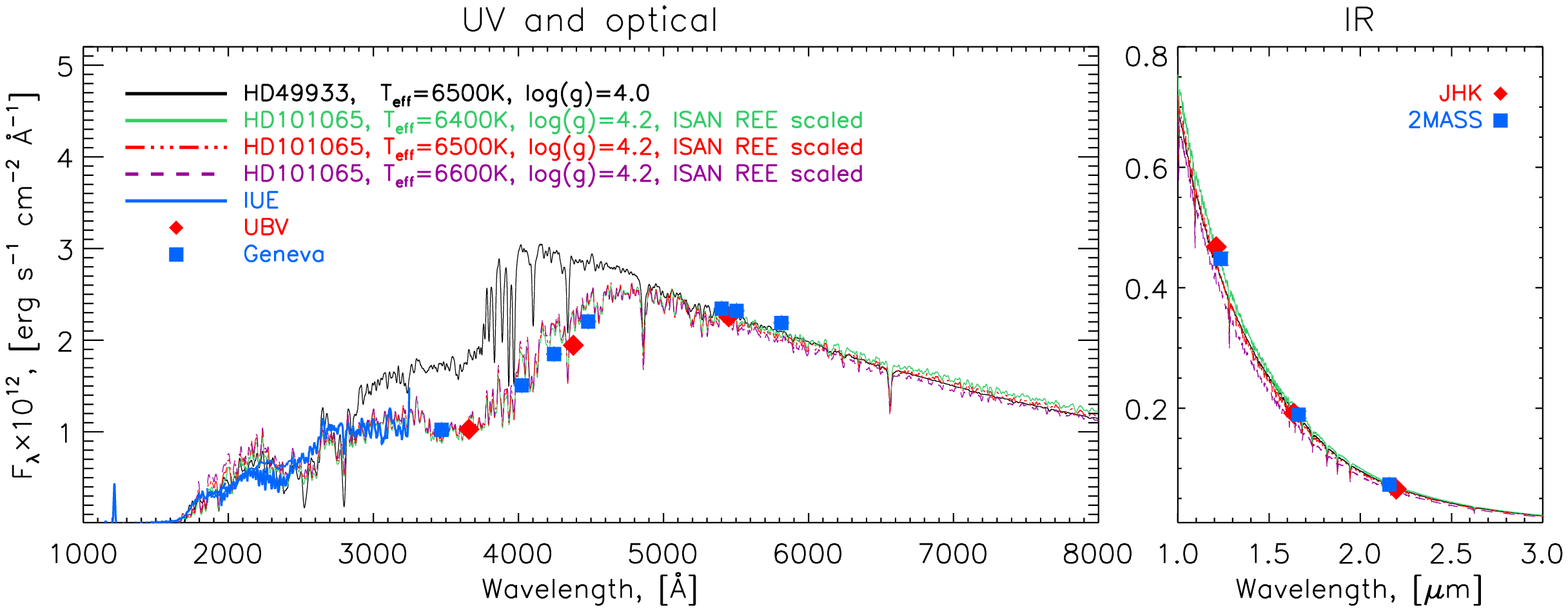}
\caption{Observed and theoretical energy distribution of \hd\ calculated with scaled REE opacity from ISAN line list (see text).
For comparison purpose the theoretical energy distribution of a normal F-type star HD~49933 is also shown. 
All theoretical calculations are smoothed with $20$\AA\ Gaussian. Theoretical fluxes were scaled assuming radiuses: 
$R=1.98R_\odot$ for HD~49933 and $\teff=6400$~K models; $R=1.90R_{\odot}$ for $\teff=6500$~K; $R=1.82R_{\odot}$ for $\teff=6600$~K.}
\label{fig:sed2}
\end{figure*}
}

The observed energy distribution of \ps\ is constructed combining the large-aperture IUE observations extracted from the INES database\footnote{{\tt http://sdc.laeff.inta.es/ines/index2.html}}
and photometric data in the $UBV$ \citep{wegner}, Geneva\footnote{{\tt http://obswww.unige.ch/gcpd/ph13.html}},
$JHK$ \citep{catalano1998} and $2MASS$ \citep{2mass} systems converted to absolute fluxes.
The model fluxes are scaled to account for the distance to \ps, which is $112\pm11$~pc according to the 
revised Hipparcos parallax of $\pi=8.93\pm0.87$~mas \citep{hipparcos}. 
Theoretical fluxes are further scaled to match the observed Paschen continuum defined by the $UBV$ and Geneva photometry. 
This scaling factor corresponds to the radius of the star, for which we found $R=1.98\pm0.03R_\odot$. The same scaling
is applied to all theoretical fluxes presented in Fig.~\ref{fig:sed}. For models with $\teff=6500$~K and $\teff=6600$~K
we find $R=1.90R_{\odot}$ and $R=1.82R_{\odot}$ respectively (see Fig.~\ref{fig:sed2}, Online Material).

It is evident that the enormous REE absorption leads to smoothing of the Balmer jump.
This is clearly visible for both models computed with the new ISAN REE line list and with the original VALD REE line list. The 
only difference is that the former one predicts less flux redward of $4100$\AA\ and more
flux in the region $3000$\AA\ and $3800$\AA\ (note that the model computed with mean REE abundances should be used with big caution
due to ignoring REE anomaly, and we present it in Fig.~\ref{fig:sed} for illustrative purpose only).
Taking into account difficulties in predicting accurately the spectra of REEs, one can conclude 
that our models computed with the recent improvements of the line REE opacity provide a reasonably good fit to the observed fluxes. They are
able to reproduce not only the shape of the observed energy distribution but also the amplitude
of the Balmer jump reasonably well, much better than in the previous attempt by ODF-based modeling \citep{piskunov2001}. However, there are still some
discrepancies with observations, i.e. too high fluxes in the 2000--2500~\AA\ region and around 4500~\AA. The description of the energy distribution of \ps\ can be probably
improved in the future once more complete line lists of REEs will become available. A new well-calibrated observed spectrophotometry in the 3300--10\,000~\AA\ region would also be extremely useful for further modeling.

The impact of the peculiar energy distribution on the photometric parameters of \hd\ is illustrated in Table~\ref{tab:colors}.
Observations in the Str\"omgren system were taken from \citet{hauck}. 
Predictions of \citet{piskunov2001} model calculated with the ODF line opacity representation
using abundances from \citep{cowley2000} are also given for comparison.
We also show theoretical photometric parameters for the models calculated with the mean REE abundances corresponding to the first and second ions,
as well as for the models with the scaled REE opacity, and for different $\teff$.
All models except the ODF-based one were computed with the chemical stratification illustrated in Fig.~\ref{fig:strat}.

There are few things to note from Table~\ref{tab:colors}. 
First of all, a strong impact of the REE opacity is clearly seen for the $c_{\rm 1}$
photometric index which reaches an even more negative value than the observed one for the models with the ISAN data and the
mean REE abundances. Similarly, the recent version of VALD already provides REE opacity that is enough to
bring down $c_{\rm 1}$ index significantly compared to previous calculations and provides a result very close
to the observed one. However, mean abundances can not be used for accurate modeling due to
significant REE ionization anomaly. Accounting for this via the scaling of REE line opacity decreases the flux redistribution
effect and thus the impact on the $c_{\rm 1}$ index. The implementation of the new ISAN data
gives a much better fit to observations than using the data from VALD alone, yet this fit is far from being fully satisfactory.

Second, it remains difficult to obtain a good fit simultaneously for all photometric indices. For instance,
all models fail to reproduce the $U-B$ value, while $b-y$ is not reproduced by models with the scaled REE opacity unless
one decreases $\teff$ to $\approx$\,6200~K. However, lowering $\teff$ is not consistent with the observed \halpha\ line.
Similar, worse fit is obtained for models with $\teff>6400$~K, which show reduced $b-y$ and 
increased $c_{\rm 1}$ indicies.

Third, stratification of Fe, Si, Ca, and Ba considered in our study does not play a significant role for the overall
energy redistribution and thus photometric colors as seen from the sixth raw of Table~\ref{tab:colors}. Most of the photometric 
indicators are negligibly affected by the presence of stratification. The $c_{\rm 1}$ index is again an exception since it
decreases by $\approx$\,0.03~mag if stratification is introduced. However, this change is still fairly small compared 
to typical precision of the observed Str\"omgren photometry.

Finally, a high sensitivity of the photometric parameters to the new ISAN calculations
indicates that the general disagreement between models and observations is the subject of future more precise and complete
calculations of REE spectra. Using the data for more REE elements than those presented in this investigation can potentially
improve the fit to the highly peculiar observed photometric properties of \ps. Implementation of the ISAN line lists for only a few REE ions 
already allowed us to achieve significantly better results
compared to previous modeling attempts and clearly showed that the ``enigmatic'' characteristics of \ps\ likely
result from the REE line opacity which was ignored in previous calculations. 
We can thus suggest that further step in the modeling of the observed properties of \ps\ should be made
in the direction of improved theoretical calculations of REE transitions.

\begin{table}
\caption{Observed and predicted colors of \hd.}
\label{tab:colors}
\begin{footnotesize}
\begin{center}
\begin{tabular}{lccccc}
\hline
\hline
                      & $b-y$            & $m_{\rm 1}$       & $c_{\rm 1}$        & $B-V$            & $U-B$\\
\hline
\textbf{Observations} & $\mathbf{0.442}$ & $\mathbf{0.427}$  & $\mathbf{-0.013}$  & $\mathbf{0.760}$ & $\mathbf{0.200}$\\
\hline
t6600g4.2, ODF        & $0.387$          & $0.582$           & $0.298$            & $0.767$          & $0.571$\\
\hline
t6400g4.2 &&&&&\\
VALD                  & $0.440$          & $0.439$           & $0.029$            & $0.745$          & $0.268$\\
REE mean abun. &&&&&\\
\hline
t6400g4.2&&&&&\\
ISAN                  & $0.456$          & $0.587$           & $-0.202$           & $0.829$          & $0.379$\\
REE mean abun.&&&&&\\
\hline
t6400g4.2&&&&&\\
VALD                  & $0.394$          & $0.420$           & $0.312$            & $0.681$          & $0.368$\\
REE scaled abun.&&&&&\\
\hline
t6400g4.2&&&&&\\
ISAN                  & $0.384$          & $0.413$           & $0.158$            & $0.663$          & $0.274$\\
REE scaled abun.&&&&&\\ 
\hline
t6500g4.2&&&&&\\
ISAN                  & $0.367$          & $0.397$           & $0.200$            & $0.634$          & $0.255$\\
REE scaled abun.&&&&&\\ 
\hline
t6600g4.2&&&&&\\
ISAN                  & $0.350$          & $0.391$           & $0.229$            & $0.607$          & $0.240$\\
REE scaled abun.&&&&&\\ 
\hline
t6400g4.2&&&&&\\
ISAN, hom.            & $0.383$          & $0.396$           & $0.188$            & $0.656$          & $0.271$\\
REE scaled abun.&&&&&\\ 
\hline
t6200g4.5&&&&&\\
ISAN                  & $0.417$          & $0.444$           & $0.024$            & $0.719$          & $0.261$\\
REE scaled abun.&&&&&\\
\hline
\end{tabular}
\end{center}
\end{footnotesize}
All models were computed with stratification of Fe, Si, Ca, and Ba, except those marked as ``ODF'' and ``hom.''
\end{table}

The strong suppression of the Balmer jump and energy redistribution to the infrared is a direct result of the unusually high
REE abundances in the stellar atmosphere. Table~\ref{tab:linestat} summarizes some statistics illustrating importance of the REE opacity in the
atmosphere of \ps. It shows the number of lines of a given ion before and after line preselection procedure
performed by the \llm\ code. This procedure selects from the master line list only those lines that noticeably
contribute to the line opacity for a given $T$-$P$ structure. Usually the preselection criterion is
$\kappa_{\rm line}/\kappa_{\rm cont}>1\%$, where $\kappa_{\rm line}$ and $\kappa_{\rm cont}$ are the line center and
continuum opacity coefficients, respectively. It is seen that, for instance, all lines of \ion{Pr}{ii} and \ion{Nd}{ii}
present in the current version of VALD contribute to the opacity in the atmosphere of \ps. A larger number of lines is selected from the more complete ISAN line lists.

\begin{table*}
\caption{Summary of REE lines in VALD and combined ISAN line lists before and after line preselection procedure.}
\label{tab:linestat}
\begin{footnotesize}
\begin{center}
\begin{tabular}{l|rrrrrrrrr}
\hline
\hline
       & \ion{Pr}{ii} & \ion{Pr}{iii} & \ion{Nd}{ii} & \ion{Nd}{iii} & \ion{Sm}{ii} & \ion{Eu}{iii} & \ion{Dy}{iii} & \ion{Ce}{ii} & \ion{Ce}{iii}\\
\hline
           \multicolumn{10}{c}{Original}\\
VALD   & $510$        & $19104$       & $1285$       & $71$          & $1346$       & $893$         & $1313$        & $16046$      & $2966$\\
ISAN   & $103428$     & $---$         & $1172579$    & $6858$        & $1047088$    & $23827$       & $31121$       & $---$        & $---$\\
\hline
\multicolumn{10}{c}{After preselection with $1$\% limit for $\teff=6400$~K, $\logg=4.2$ model, REE scaled abundances}\\
VALD   & $510$        & $6539$        & $1285$       & $71$          & $1121$       & $58$          & $1134$        & $15908$      & $1270$\\
ISAN   & $83167$      & $---$         & $74413$      & $3810$	     & $58637$      & $74$	    & $21715$	    & $---$	   & $---$\\
\hline
\end{tabular}
\end{center}
\end{footnotesize}
\end{table*}

\begin{figure*}
\includegraphics[width=4.cm,angle=-90]{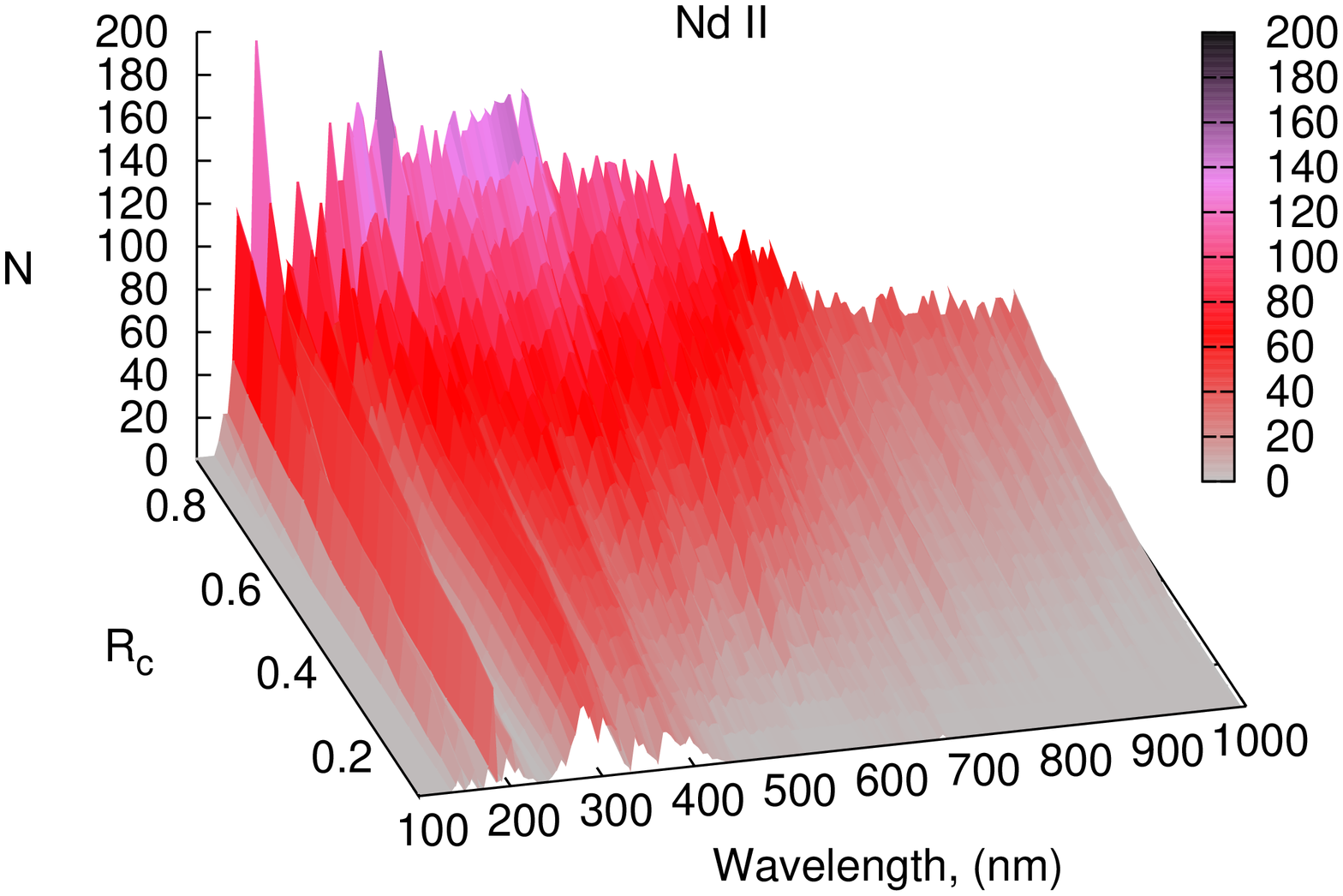}
\includegraphics[width=4.cm,angle=-90]{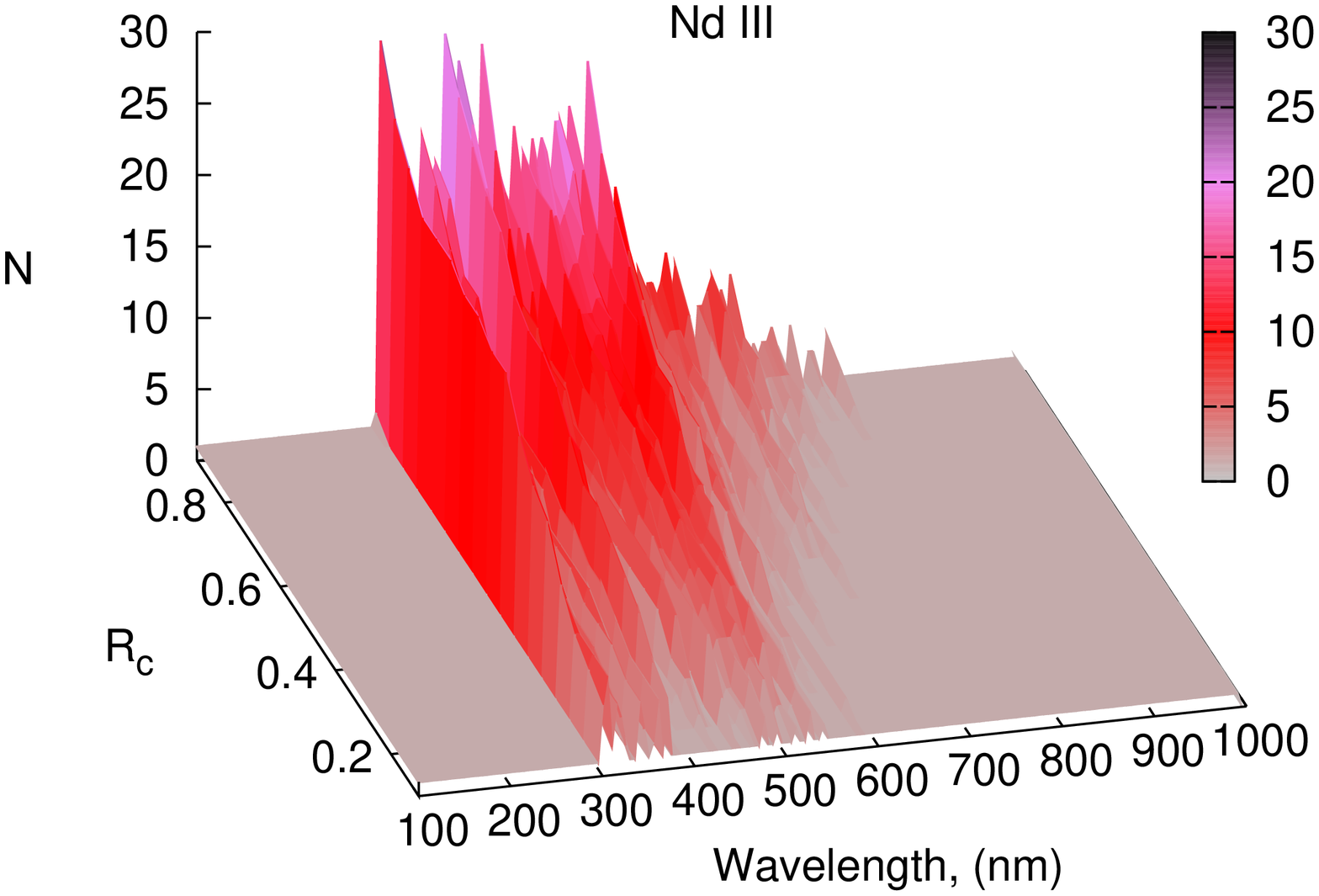}
\includegraphics[width=4.cm,angle=-90]{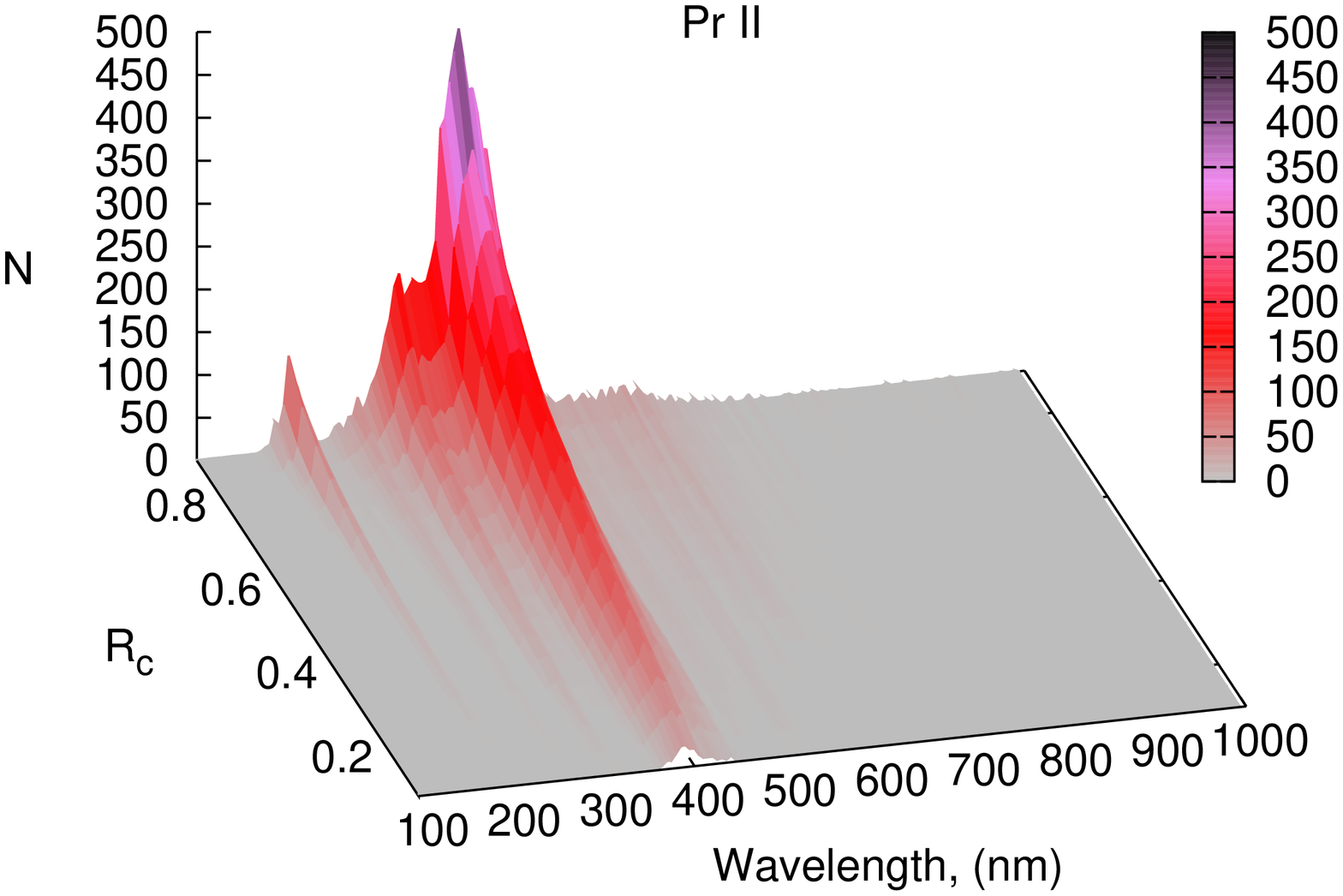}\vspace{0.5cm}
\includegraphics[width=4.cm,angle=-90]{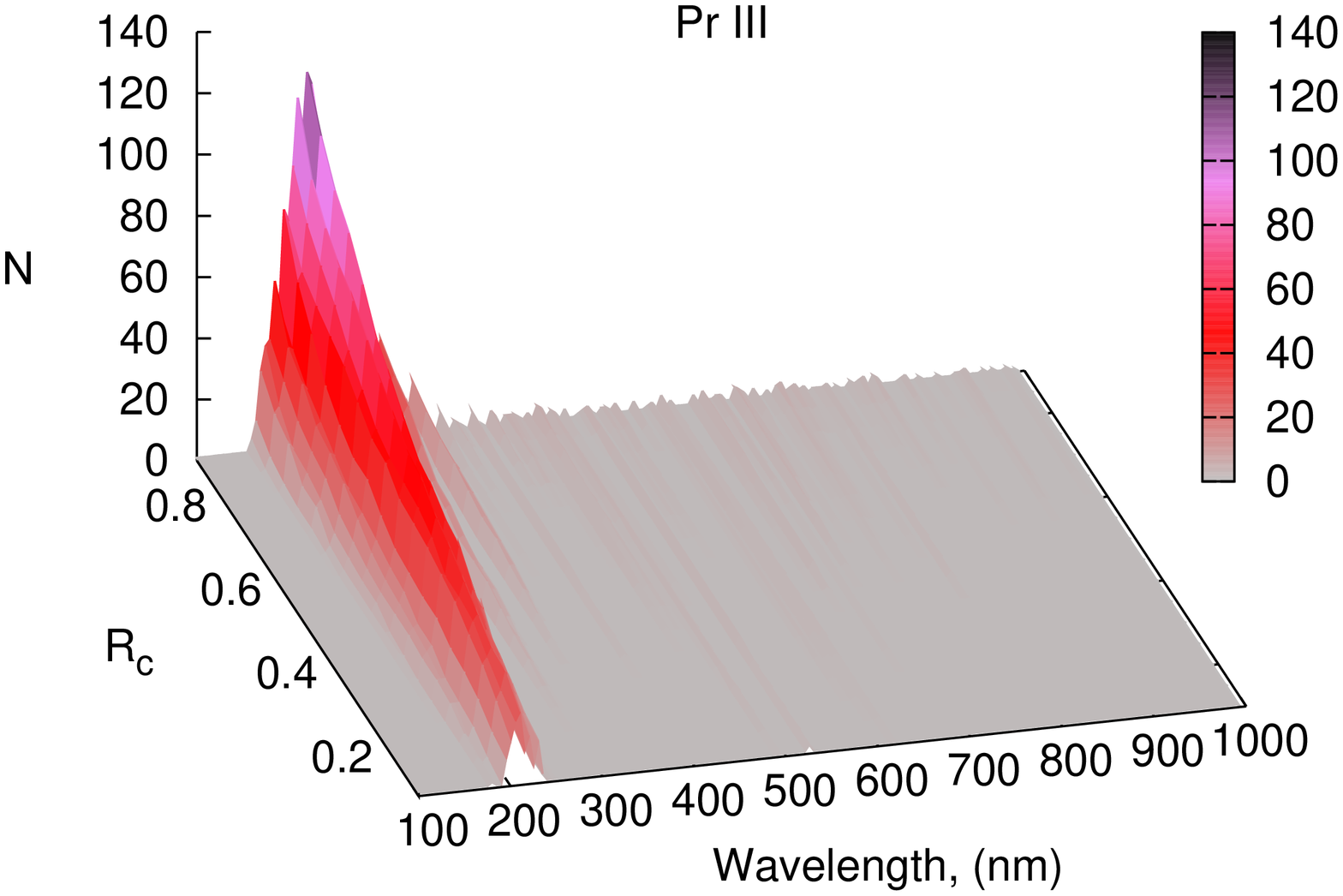}
\includegraphics[width=4.cm,angle=-90]{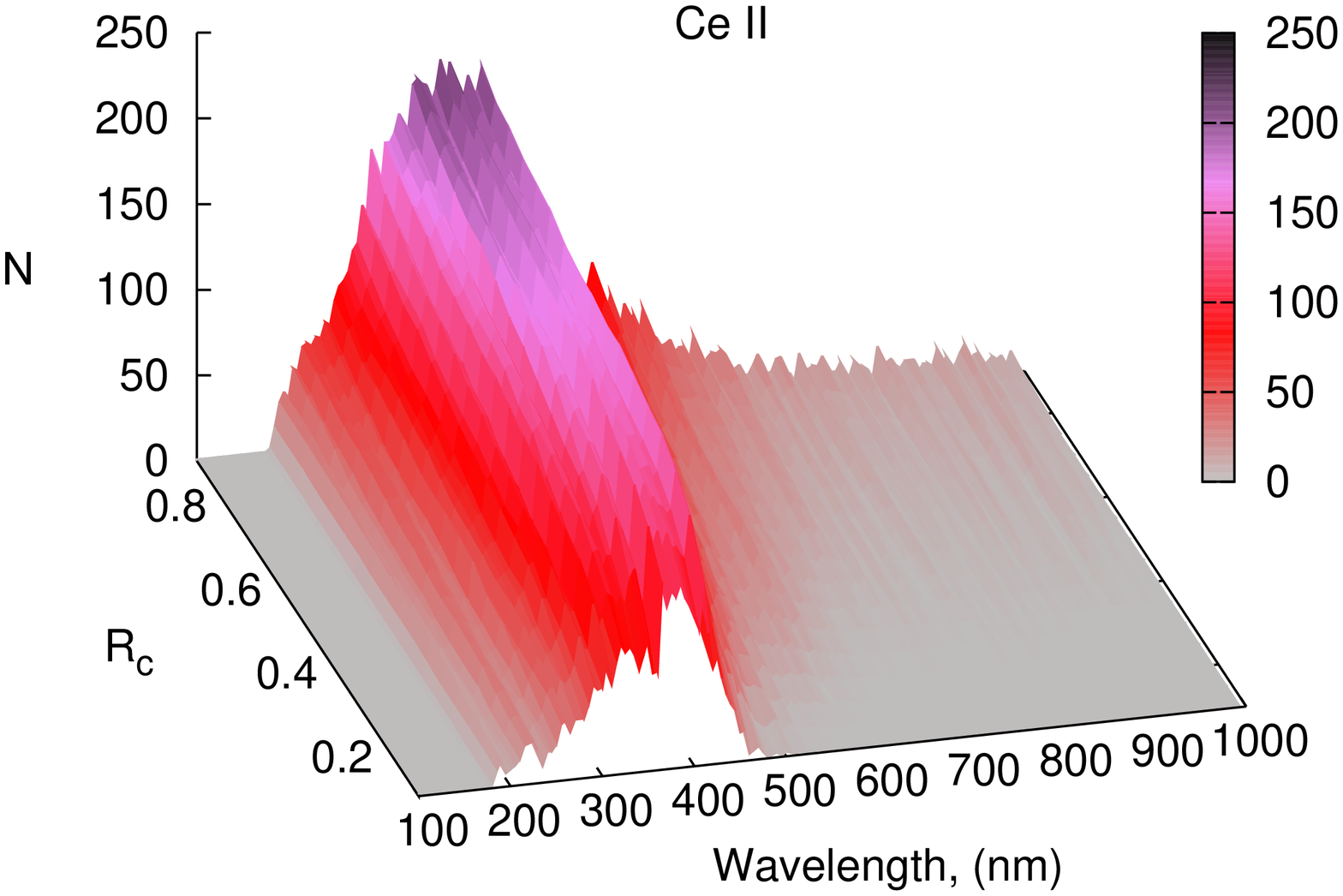}
\includegraphics[width=4.cm,angle=-90]{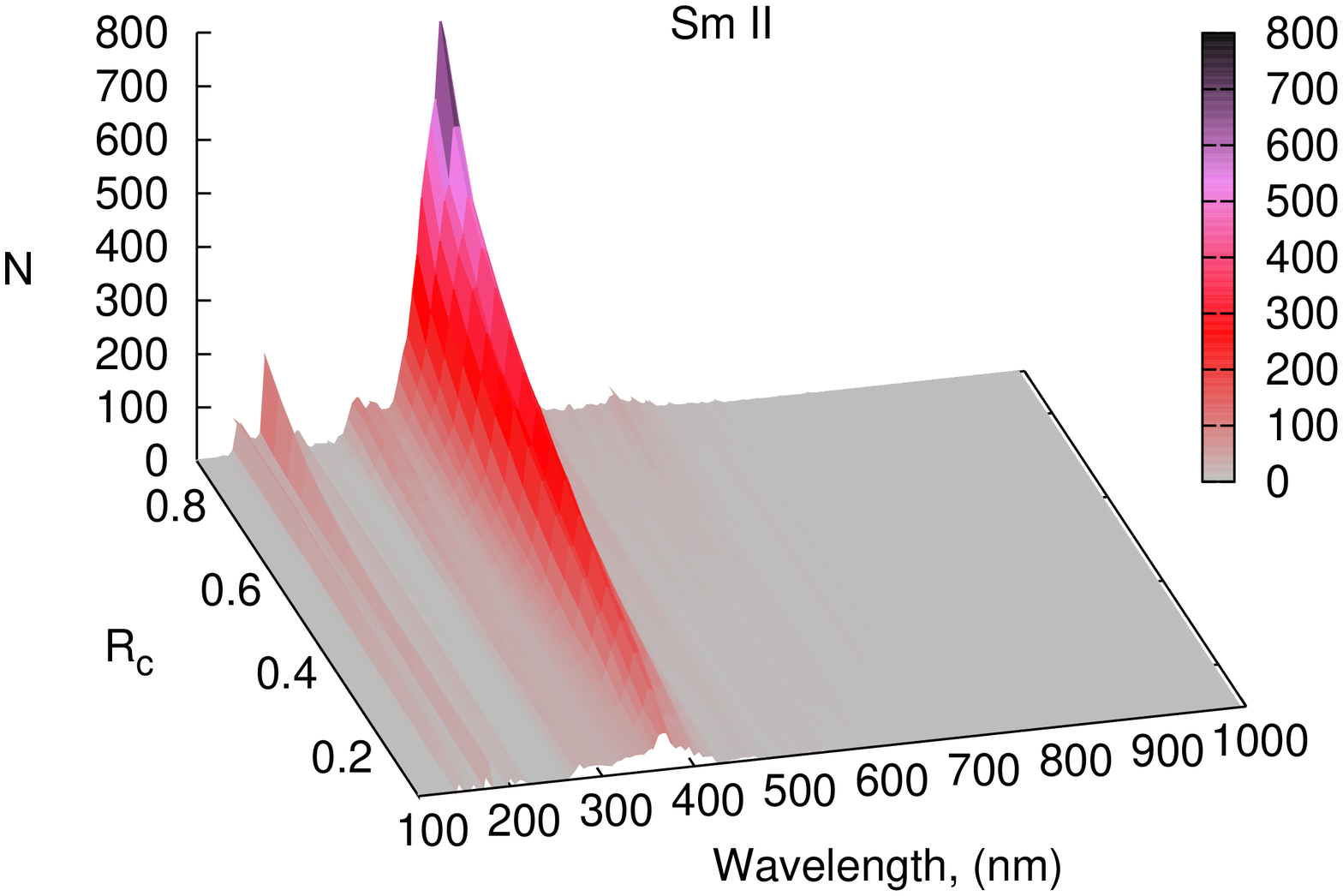}
\caption{Distribution of lines for several REE ions depending on their position and central depth in the ISAN line
lists for the model atmosphere with $\teff=6400$~K, $\logg=4.2$.}
\label{fig:linedist}
\end{figure*}

As expected from the low magnitude of the $c_{\rm 1}$ index and the energy distribution shown in 
Fig.~\ref{fig:sed}, a great majority of strong REE lines are concentrated around the
Balmer jump. Figure~\ref{fig:linedist} illustrates the line distribution for selected REE ions depending upon 
their wavelength position and central intensity. We show the number of lines per 50~\AA\ wavelength bin for
which the central intensities are greater or equal to a given normalized intensity. The latter ranges from $0$ 
(fully saturated strong line) to $1$ (continuum level, weak line).
One can seen that ions like \ion{Ce}{ii}, \ion{Pr}{ii},
\ion{Nd}{ii/iii}, \ion{Sm}{ii} have a large number of strong lines located exactly in the region of Balmer jump and
blueward, dramatically contributing to the opacity and radiation field.

\section{Conclusions}
\label{sec:conclusions}
\hd\ is known to be one of the most peculiar stars ever found. Its unusual photometric colors and extremely
rich spectra of REE elements represent a major challenge for modern
stellar atmosphere theory. With the increasing number of the accurately
measured and calculated REE transitions and the progress in the model atmosphere techniques it became possible to carry out this quantitatively 
new analysis of the spectrum of \hd\ and to re-determine atmospheric parameters of this star.

In this study we investigated the impact of the new theoretically computed spectra of some REE elements 
on the atmospheric properties of \hd\ and carried out a detailed spectroscopic analysis of the stratification 
of Si, Fe, Ba, and Ca in the stellar atmosphere.
Theoretical computations of REE transitions have extended already existing line lists of REE lines by a factor of one thousand.
Implementing an iterative procedure of the abundance analysis, we re-derived atmospheric parameters of the star
in a self-consistent way, accounting for the ionization disequilibrium between the first and second ions of the REEs
caused by significant NLTE effects.

The main results of our investigation can be summarized in the following:
\begin{itemize}
\item
REEs appear to play a key role in the radiative energy balance in the atmosphere of \hd, leading
to the strong suppression of the Balmer jump and energy redistribution from UV and visual to IR.
\item
The abnormal photometric color-indices of \hd\ arise due to a strong absorption in REE lines. 
We showed that the introduction of the new extensive line lists of REEs allow one to achieve a better 
agreement between the unusually low magnitude of the Str\"omgren $c_{\rm 1}$ index observed for this star
and the model predictions. We also demonstrated, for the first time, a satisfactory agreement between the observed and computed spectral energy distribution of \hd.
\item
The remaining discrepancy between the observed and calculated photometric parameters indicates that some important
opacity sources are still missing in the modeling process. For this reason the future extension of the REE line lists
is needed for complete understanding of the observed properties of \hd.
\item
Using theoretical spectrum synthesis we derived stratification profiles for Si, Ca, Fe, and Ba. We found
a strong depletion of all these elements in the upper atmosphere of \hd. The effect of stratification
on the \halpha\ line profile appears to be at the level of $1$\%, which is however important for accurate
calculation of the hydrogen line profiles.
\item
There is not much influence of the stratification on the considered photometric parameters of \hd, except for the Str\"omregn $c_{\rm 1}$ index, which is modified by stratification by $0.03$~mag.
\item
Using combined photometric and spectroscopic analysis and based on our novel iterative procedure of
the abundance and stratification analysis we find effective temperature of the \hd\ to be $\teff=6400$~K.
\end{itemize}

Taking into account the current progress in the atmospheric modeling, understanding of the REE opacity,
and detailed spectroscopic analysis we conclude that there is nothing special in \hd\ compared to other known cool Ap and 
roAp stars except its relatively low $\teff$ and a high abundance of REEs. Consequently, for this moment \hd\ appears 
to be the coolest Ap and roAp star known but otherwise is a typical member of these groups.

\section{Discussion}
\label{sec:discussion}

\begin{figure}
\includegraphics[width=\hsize]{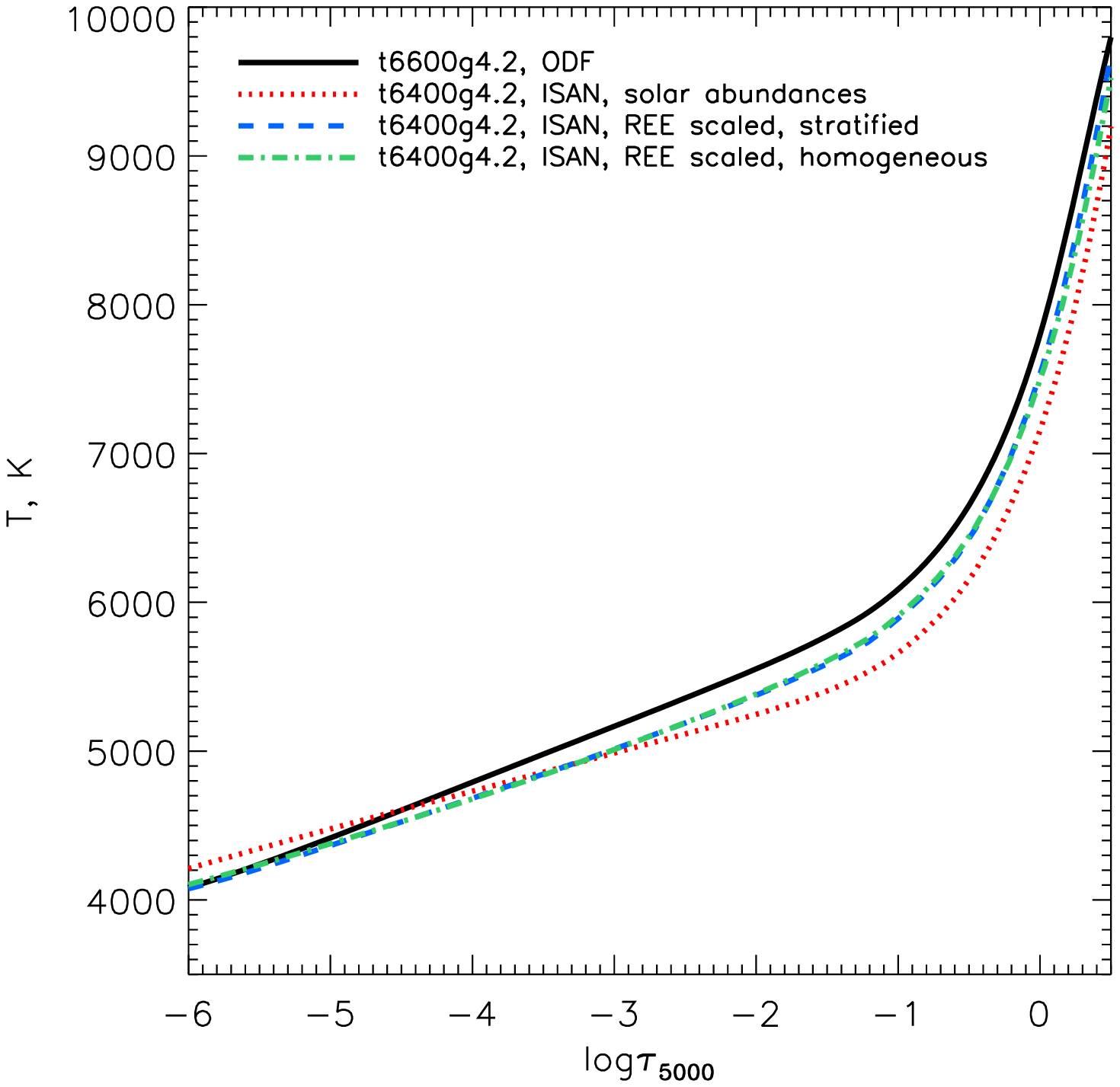}
\caption{Comparison in atmospheric temperature structure of theoretical models. Thick line~--~ ODF model from \citet{piskunov2001},
dotted line~--~homogeneos solar abundances model, dashed line~--~stratified abundances final model adopted in the present study,
dash-dotted line~--~same as before but assuming homogeneos abundances.}
\label{fig:ttau}
\end{figure}

We can compare the global parameters of \hd\, derived in the present study with the values obtained by an independent pulsation modeling \citep{puls08}. The best model explaining the observed frequency pattern
of \hd\, is characterized by $\teff$\,=\,6622$\pm$100 K, $\logg$\,=\,4.06$\pm$0.04, R\,=\,$1.90\pm0.08$\,$R_\odot$. These parameters
agree with ours within $2\sigma$. However, to fit the observed frequencies in the framework of current asterosismic models one needs dipolar magnetic field strength $B_{\rm p}$\,=\,8.7~kG, which is incompatible with the observed field modulus of $\langle B \rangle$\,=\,2.3~kG.
One of the reasons for this significant discrepancy between the observed and inferred magnetic field is
the use of normal $T$-$\tau$ relation in the pulsation modeling. This relation
differs significantly from a more realistic one derived in the present paper, as illustrated in Fig.~\ref{fig:ttau} 
(Note that computing model with solar abundances we did not apply scaling of the REE opacity due to the low abundance of REE elements
in solar atmosphere and thus its negligible impact on model structure). As expected from the fit to \halpha\ line and
photometric parameters presented above, the effect of stratification on model $T-P$ structure is negligible and hardly
distinguishable from the homogeneous abundance model (dash-dotted line in Fig.~\ref{fig:ttau}). Note that it is still
considereably different from the solar abundance model, and is in good agreement with the result of
\citet{monin2007}.

In spite of our new results, illustrating an improved understanding of the physics of the atmosphere of \hd, 
it is still impossible to describe all observed properties of this star with the current state-of-the-art modeling. 
In this section we would like to discuss physical effects that we ignored in our analysis.

Taking into account a jungle of the REE lines seen in the spectrum of \hd\ and their importance as an opacity source for the model atmosphere calculation it would be reasonable to search for the possible
stratification of rare-earth elements. Considering the case of a hotter roAp star HD~24712, for which the accumulation
of REEs in the upper atmosphere plays a critical role, producing a characteristic inverse temperature gradient 
\citep[see][for more details]{hd24712}, one would expect the same mechanism to operate in the atmosphere of \hd\ as well.
However, the stratification analysis of REEs is highly complicated due to blending of REE lines with accurately known
atomic parameters (that ideally should be used for such an analysis) by other REE lines for that no laboratory measurements
exist and thus can not be included in the spectrum synthesis. At the same time, 
assuming that REEs are distributed similarly in both stars,
we do not expect a strong impact of the REE stratification
on the energy distribution of \hd\ simply because it would affect only the optically thin atmospheric layers.

As mentioned above, due to the low $\teff$ of \ps, it is impossible to infer the $\logg$ value from the hydrogen
lines. Furthermore, such $\logg$ indicator as the $c_{\rm 1}$ photometric index is strongly affected by the REE opacity, which is
still far from being completely understood. The ISAN calculations gave us information on several REE ions but the role
of other REE species is not known at this moment. On the other hand, as seen from the high resolution spectra of \hd, most of the 
line absorption is due to such ions as \ion{Pr}{ii/iii}, \ion{Nd}{ii/iii}, and \ion{Ce}{ii/iii} that are represented in
our calculations by a large number of lines. This suggests that the magnitude of the energy redistribution is well
modeled. However, extending our calculations using new line lists for other REE ions will allow to improve the accuracy of theoretical $c_{\rm 1}$ index and constrain the $\logg$ of the star. In principle,
increasing $\logg$ up to 4.5 and decreasing $\teff$ down to 6200~K provides a better fit to the observed photometric
parameters as demonstrated in the last row of Table~\ref{tab:colors}. Also the fit to the \halpha\ line remains reasonable with these parameters. However,
the ionization equilibrium for different elements is not as good as with the $\teff=6400$~K, $\logg=4.2$ model. This is why, and taking into account various uncertainties related to the REE opacity, we did not attempt to derive surface gravity in our study.
This limitation does not affect any of the results presented in this paper.

\begin{acknowledgements}
This work was supported by the following grants: FWF Lise Meitner grant Nr. M998-N16 and Deutsche Forschungsgemeinschaft (DFG)
Research Grant RE1664/7-1 to DS, by RFBG grants (08-02-00469a, 09-02-00002a), Presidium RAS Programme ``Origin and evolution of stars and galaxies''
and Russian Federal Agency on Science and Innovation Programme (02.740.11.0247) to TR.  
OK is a Royal Swedish Academy of Sciences Research Fellow supported by grants from the Knut and Alice Wallenberg Foundation and the Swedish Research Council.
This publication makes use of data products from the Two Micron All Sky Survey, which is a joint project 
of the University of Massachusetts and the Infrared Processing and Analysis Center/California 
Institute of Technology, funded by the National Aeronautics and Space Administration 
and the National Science Foundation. Based on INES data from the IUE satellite.
We also acknowledge the use of electronic databases (VALD, SIMBAD, NIST, NASA's ADS)
\end{acknowledgements}

\Online
%

\begin{table*}
\caption{List of Si, Ca, Fe, and Ba spectral lines used for reconstruction of
chemical stratification.}
\begin{footnotesize}
\begin{center}
\begin{tabular}{lcrrrl}
\noalign{\smallskip}
\hline
\hline
Ion &$\lambda$ (\AA) &\ei\,(eV)  &\loggf&$\log\,\gamma_{\rm St}$ & Ref.\\
\hline
\ion{Si}{i} &  5517.533&  5.082 &-2.49~&  -4.46 &solar\\  
\ion{Si}{i} &  5690.425&  4.930 &-1.76~&  -4.57 &NIST \\            
\ion{Si}{i} &  5797.688&  5.614 &-2.67~&  -3.86 &K07  \\           
\ion{Si}{i} &  5797.856&  4.954 &-2.05~&  -4.32 &NIST \\        
\ion{Si}{i} &  6087.697&  5.871 &-2.38~& approx &K07  \\        
\ion{Si}{i} &  6087.805&  5.871 &-1.71~& approx &solar\\         
\ion{Si}{i} &  6155.134&  5.619 &-0.80~&  -3.70 &solar\\         
\ion{Si}{ii}&  6347.109&  8.121 &~0.297&  -5.04 &BBCB \\         
\ion{Si}{ii}&  6371.371&  8.121 &-0.003&  -5.04 &BBCB \\
\ion{Si}{i} &  8742.446&  5.871 &-0.510&  -4.52 &solar\\         
\ion{Si}{i} &  8892.720&  5.984 &-0.830&  -4.37 &solar\\
            &          &        &      &        &     \\         
\ion{Ca}{i} &  6717.681&  2.709 &-0.524&  -4.90 &SR   \\            
\ion{Ca}{i} &  7148.150&  2.709 &~0.137&  -6.01 &SR   \\          
\ion{Ca}{i} &  7326.145&  2.933 &-0.208&  -5.16 &S    \\         
\ion{Ca}{ii}&  8254.721&  7.515 &-0.398&  -4.60 &T    \\         
\ion{Ca}{ii}&  8498.223&  1.692 &-1.416&  -5.70 &T    \\         
\ion{Ca}{ii}&  8912.068&  7.047 &~0.637&  -5.10 &T    \\        
            &          &        &      &        &     \\         
\ion{Fe}{i} &  4404.750&	 1.557 &-0.142&  -6.20 &MFW  \\         
\ion{Fe}{ii}&  5169.033&	 2.891 &-1.14~&  -6.59 &M    \\   
\ion{Fe}{i} &  5405.775&	 0.990 &-1.844&  -6.30 &MFW  \\  
\ion{Fe}{i} &  5415.199&	 4.386 &~0.642&  -4.76 &BWL  \\  
\ion{Fe}{i} &  5615.644&	 3.332 &~0.050&  -5.50 &BKK  \\  
\ion{Fe}{i} &  6219.281&	 2.198 &-2.433&  -6.20 &MFW  \\  
\ion{Fe}{ii}&  7222.394&	 3.889 &-3.36~&  -6.67 &HLGH \\  
\ion{Fe}{i} &  8248.129&	 4.371 &-0.887&  -5.36 &K07  \\          	  
\ion{Fe}{i} &  8365.634&	 3.251 &-2.047&  -6.10 &BK   \\        			  
\ion{Fe}{i} &  8945.189&	 5.033 &-0.235&  -5.17 &K07  \\        		
            &          &        &      &        &     \\
\ion{Ba}{i} &  5777.618&	 1.676 &~0.72~&  approx&MW   \\         
\ion{Ba}{ii}&  5853.668&	 0.604 &-1.00~&  -6.35 &MW   \\         
\ion{Ba}{i} &  6019.465&	 1.120 &-0.10~&  approx&MW   \\           
\ion{Ba}{i} &  6063.109&	 1.143 &~0.21~&  approx&MW   \\  
\ion{Ba}{ii}&  6141.713&	 0.704 &-0.076&  -6.35 &MW   \\         
\ion{Ba}{ii}&  6496.897&	 0.604 &-0.377&  -6.35 &MW   \\         
\ion{Ba}{i} &  7059.938&	 1.190 &~0.68~&  approx&MW   \\  
\ion{Ba}{i} &  7280.297&	 1.143 &~0.47~&  approx&MW   \\  
\hline                                
\end{tabular}
\label{tab:strat}
\end{center}
Columns give the ion identification, central wavelength, excitation
potential, oscillator strength (\loggf) and the Stark damping constant
($\log\,\gamma_{\rm St}$) for $T=10\,000$\,K. The last column gives reference for the adopted oscillator strength:

NIST~--~\citet{NIST}; K07~--~\citet{K07}; BBCB~--~\citet{BBCB};  
SR~--~\citet{SR}; S~--~\citet{S}; T~--~\citet{T}; 
MFW~--~\citet{MFW}; BWL~--~\citet{BWL}; BKK~--~\citet{BKK}; HLGH~--~\citet{HLGH};
BK~--~\citet{BK}; MW~--~\citet{MW}.
\end{footnotesize}
\end{table*}

\begin{table*}
\caption{Abundances of individual elements.}
\begin{scriptsize}
\begin{center}
\begin{tabular}{lcccc}
\hline\hline
Ion & $N_{\rm lines}$ & $\log(N_{\rm el}/N_{\rm total})$ & Error & $\log(N_{\rm el}/N_{\rm total})_{\bigodot}$\\
\hline                                                          
\ion{C}{i}    & $4 $  &  $-3.91$  &  $\pm 0.20$  &  $-3.65$\\   
\ion{O}{i}    & $2 $  &  $-3.42$  &  $\pm 0.36$  &  $-3.38$\\   
\ion{Na}{i}   & $4 $  &  $-6.03$  &  $\pm 0.22$  &  $-5.87$\\
\ion{Mg}{ii}  & $1 $  &  $-4.41$  &  $        $  &  $-4.51$\\   
\ion{Al}{i}   & $1 $  &  $-6.69$  &  $        $  &  $-5.67$\\   
\ion{S}{i}    & $1 $  &  $-4.76$  &  $        $  &  $-4.90$\\   
\ion{Sc}{ii}  & $6 $  &  $-8.97$  &  $\pm 0.59$  &  $-8.87$\\   
\ion{Ti}{i}   & $6 $  &  $-7.38$  &  $\pm 0.31$  &  $-7.14$\\
\ion{Ti}{ii}  & $12$  &  $-7.32$  &  $\pm 0.30$  &  $-7.14$\\        
\ion{V}{i}    & $8 $  &  $-7.15$  &  $\pm 0.41$  &  $-8.04$\\
\ion{V}{ii}   & $6 $  &  $-7.28$  &  $\pm 0.32$  &  $-8.04$\\        
\ion{Cr}{i}   & $12$  &  $-6.45$  &  $\pm 0.34$  &  $-6.40$\\   
\ion{Cr}{ii}  & $16$  &  $-5.91$  &  $\pm 0.30$  &  $-6.40$\\        
\ion{Mn}{i}   & $8 $  &  $-6.26$  &  $\pm 0.29$  &  $-6.65$\\   
\ion{Mn}{ii}  & $3 $  &  $-5.68$  &  $\pm 0.20$  &  $-6.65$\\        
\ion{Co}{i}   & $40$  &  $-6.00$  &  $\pm 0.22$  &  $-7.12$\\
\ion{Co}{ii}  & $2 $  &  $-5.09$  &  $\pm 0.47$  &  $-7.12$\\        
\ion{Ni}{i}   & $8 $  &  $-7.01$  &  $\pm 0.47$  &  $-5.81$\\
\ion{Cu}{i}   & $2 $  &  $-7.97$  &  $\pm 0.05$  &  $-7.83$\\
\ion{Sr}{i}   & $1 $  &  $-7.60$  &  $        $  &  $-9.12$\\
\ion{Sr}{ii}  & $2 $  &  $-7.91$  &  $\pm 0.07$  &  $-9.12$\\        
\ion{Y}{i}    & $1 $  &  $-7.97$  &  $        $  &  $-9.83$\\
\ion{Y}{ii}   & $8 $  &  $-9.17$  &  $\pm 0.24$  &  $-9.83$\\        
\ion{Zr}{i}   & $5 $  &  $-7.90$  &  $\pm 0.46$  &  $-9.45$\\   
\ion{Zr}{ii}  & $11$  &  $-7.53$  &  $\pm 0.21$  &  $-9.45$\\
\ion{Nb}{i}   & $4 $  &  $-7.48$  &  $\pm 0.25$  &  $-10.62$\\  
\ion{Mo}{i}   & $5 $  &  $-7.56$  &  $\pm 0.29$  &  $-10.12$\\
\ion{Ru}{i}   & $4 $  &  $-7.57$  &  $\pm 0.34$  &  $-10.20$\\
\ion{Rh}{i}   & $2 $  &  $-7.16$  &  $\pm 0.61$  &  $-10.92$\\
\ion{Pd}{i}   & $6 $  &  $-6.98$  &  $\pm 0.23$  &  $-10.38$\\
\ion{Cd}{i}   & $1 $  &  $-7.94$  &  $        $  &  $-10.27$\\
\ion{Sn}{ii}  & $1 $  &  $-7.48$  &  $        $  &  $-10.04$\\
\ion{La}{ii}  & $31$  &  $-7.91$  &  $\pm 0.36$  &  $-10.91$\\  
\ion{Ce}{ii}  & $118$ &  $-7.53$  &  $\pm 0.32$  &  $-10.34$\\ 
\ion{Ce}{iii} & $4 $  &  $-5.70$  &  $\pm 0.24$  &  $-10.34$\\        
\ion{Pr}{ii}  & $37$  &  $-8.54$  &  $\pm 0.31$  &  $-11.46$\\        
\ion{Pr}{iii} & $26$  &  $-6.58$  &  $\pm 0.38$  &  $-11.46$\\        
\ion{Nd}{i}   & $4 $  &  $-6.88$  &  $\pm 0.12$  &  $-10.59$\\
\ion{Nd}{ii}  & $109$ &  $-7.11$  &  $\pm 0.34$  &  $-10.59$\\       
\ion{Nd}{iii} & $8 $  &  $-5.97$  &  $\pm 0.60$  &  $-10.59$\\        
\ion{Sm}{i}   & $1 $  &  $-7.11$  &  $        $  &  $-11.04$\\
\ion{Sm}{ii}  & $63$  &  $-7.25$  &  $\pm 0.39$  &  $-11.04$\\        
\ion{Eu}{ii}  & $7 $  &  $-8.08$  &  $\pm 0.45$  &  $-11.52$\\  
\ion{Gd}{ii}  & $59$  &  $-7.34$  &  $\pm 0.33$  &  $-10.92$\\        
\ion{Tb}{ii}  & $3 $  &  $-8.84$  &  $\pm 0.22$  &  $-11.76$\\  
\ion{Tb}{iii} & $10$  &  $-6.07$  &  $\pm 0.43$  &  $-11.76$\\        
\ion{Dy}{i}   & $1 $  &  $-7.88$  &  $        $  &  $-10.90$\\
\ion{Dy}{ii}  & $29$  &  $-7.45$  &  $\pm 0.33$  &  $-10.90$\\        
\ion{Dy}{iii} & $9 $  &  $-5.78$  &  $\pm 0.33$  &  $-10.90$\\        
\ion{Ho}{i}   & $1 $  &  $-6.79$  &  $        $  &  $-11.53$\\
\ion{Ho}{iii} & $4 $  &  $-6.34$  &  $\pm 0.11$  &  $-11.53$\\        
\ion{Er}{i}   & $1 $  &  $-8.16$  &  $        $  &  $-11.11$\\
\ion{Er}{ii}  & $23$  &  $-7.81$  &  $\pm 0.30$  &  $-11.11$\\        
\ion{Er}{iii} & $4 $  &  $-6.18$  &  $\pm 0.29$  &  $-11.11$\\        
\ion{Tm}{ii}  & $18$  &  $-8.16$  &  $\pm 0.32$  &  $-12.04$\\  
\ion{Yb}{i}   & $1 $  &  $-7.44$  &  $\       $  &  $-10.96$\\
\ion{Yb}{ii}  & $9 $  &  $-8.98$  &  $\pm 0.37$  &  $-10.96$\\
\ion{Lu}{ii}  & $7 $  &  $-8.44$  &  $\pm 0.30$  &  $-11.98$\\  
\ion{Hf}{i}   & $1 $  &  $-7.89$  &  $        $  &  $-11.16$\\
\ion{Hf}{ii}  & $3 $  &  $-8.39$  &  $\pm 0.30$  &  $-11.16$\\
\ion{Ta}{i}   & $3 $  &  $-8.59$  &  $\pm 0.49$  &  $-12.21$\\
\ion{Ta}{ii}  & $1 $  &  $-8.32$  &  $        $  &  $-12.21$\\
\ion{W}{i}    & $4 $  &  $-7.83$  &  $\pm 0.61$  &  $-10.93$\\
\ion{W}{ii}   & $3 $  &  $-7.79$  &  $\pm 0.35$  &  $-10.93$\\
\ion{Re}{i}   & $2 $  &  $-7.60$  &  $\pm 0.04$  &  $-11.81$\\
\ion{Ir}{i}   & $2 $  &  $-7.17$  &  $\pm 0.11$  &  $-10.66$\\
\ion{Pt}{i}   & $2 $  &  $-7.43$  &  $\pm 0.12$  &  $-10.40$\\
\ion{Au}{i}   & $1 $  &  $-8.33$  &  $\pm 0.00$  &  $-11.03$\\
\ion{Hg}{i}   & $2 $  &  $-7.76$  &  $\pm 0.15$  &  $-10.91$\\
\ion{Th}{ii}  & $9 $  &  $-9.18$  &  $\pm 0.35$  &  $-11.98$\\
\ion{Th}{iii} & $5 $  &  $-8.09$  &  $\pm 0.18$  &  $-11.98$\\
\ion{U}{ii}   & $17$  &  $-8.44$  &  $\pm 0.15$  &  $-12.56$\\
\hline
\end{tabular}
\label{tab:abn}
\end{center}
\end{scriptsize}
\smallskip
The solar abundances are taken from \citet{solabn}.
\end{table*}

\end{document}